\documentclass[%
 reprint,
 amsmath,amssymb,
 aps,
]{revtex4-2}
\usepackage{graphicx}
\usepackage{dcolumn}
\usepackage{bm}

\usepackage[T1]{fontenc}
\usepackage[utf8]{inputenc}

\usepackage{siunitx} 

\usepackage{layout} 

\usepackage{hyperref}
\hypersetup{
      filecolor=blue,
      urlcolor=blue
      colorlinks=true,
      linkcolor=blue,
      citecolor=blue}

\usepackage{graphicx}
\usepackage{subfig}
\usepackage{physics}
\usepackage{amssymb}
\usepackage{amsmath}

\usepackage[dvipsnames]{xcolor}
\usepackage{cuted}
\usepackage[skip = 2pt, font=footnotesize]{caption}
\usepackage{soul}
\usepackage{mathtools}
\usepackage{lineno}

\begin{document}
\preprint{APS/123-QED}

\title{Mitigation of exchange crosstalk in dense quantum dot arrays}%

\author{Daniel Jirovec$^1$}
\thanks{These authors contributed equally to this work}
\email{d.jirovec@tudelft.nl}

\author{Pablo Cova Fari\~na$^{1*}$}

\author{Stefano Reale$^1$}
\author{Stefan D. Oosterhout$^{1,2}$}
\author{Xin Zhang$^1$}
\author{Sander de Snoo$^1$}
\author{Amir Sammak$^{2}$}
\author{Giordano Scappucci$^1$}
\author{Menno Veldhorst$^1$}
\author{Lieven M. K. Vandersypen$^1$}
\email{L.M.K.Vandersypen@tudelft.nl}

\affiliation{
$^1$QuTech and Kavli Institute of Nanoscience, Delft University of Technology, 2600 GA Delft, The Netherlands\\
$^2$ Netherlands Organisation for Applied Scientific Research (TNO), 2628 CK Delft, The Netherlands\\}
\date{\today}

\begin{abstract}
Coupled spins in semiconductor quantum dots are a versatile platform for quantum computing and simulations of complex many-body phenomena. 
However, on the path of scale-up, cross-talk from densely packed electrodes poses a severe challenge. While cross-talk onto the dot potentials is nowadays routinely compensated for, cross-talk on the exchange interaction is much more difficult to tackle because it is not always directly measurable. Here we propose and implement a way of characterizing and compensating cross-talk on adjacent exchange interactions by following the singlet-triplet avoided crossing in Ge. We show that we can easily identify the barrier-to-barrier cross-talk element without knowledge of the particular exchange value in a 2$\times$4 quantum dot array. We uncover striking differences among these cross-talk elements which can be linked to the geometry of the device and the barrier gate fan-out. We validate the methodology by tuning up four-spin Heisenberg chains. The same methodology should be applicable to longer chains of spins and to other semiconductor platforms in which mixing of the singlet and the lowest-energy triplet is present or can be engineered.   
Additionally, this procedure is well suited for automated tuning routines as we obtain a stand-out feature that can be easily tracked and directly returns the magnitude of the cross-talk. 

\end{abstract}
\maketitle
\section{Introduction}
Spin qubits in gate defined semiconductor quantum dots constitute a versatile platform for quantum computation owing to their long coherence times, demonstrated high fidelity single- and two-qubit gates as well as their small footprint~\cite{Stano2022, Burkard2023}. Also, they find applications in quantum simulations due to the inherent tunability of most Hamiltonian parameters which allows to explore different limits of the Fermi-Hubbard and the Heisenberg model~\cite{Barthelemy2013}.
A challenge in scaling up, however, is cross-talk from the gates defining the potential landscape as sketched in Fig. \ref{fig1}a. Several approaches for cross-talk management exist and rely on defining a set of virtual gates designed to control the energy scale of choice, be it the on-site potential, tunnel coupling, or the exchange interaction~\cite{Volk2019, Diepen2018, Qiao2020, Mills2019a, Rao2025}. Virtual plunger gates controlling the on-site potential are nowadays routinely used in experiments (Fig. \ref{fig1}b), but methods for barrier-to-barrier cross-talk compensation are typically overlooked (Fig. \ref{fig1}c) and only rarely implemented~\cite{Fedele2021, Madzik2025}. 

Digital spin qubit experiments so far circumvent the problem by avoiding the simultaneous activation of adjacent exchange couplings~\cite{Takeda2021, Philips2022} or by populating quantum dot arrays only sparsely~\cite{Vandersypen2017, Boter2022, Wang2024}. When only one exchange interaction is activated at a time, any cross-talk to other exchanges will not induce a detrimental effect, because the exponential dependence of exchange strength on barrier voltage leads to a wide voltage range over which the other interactions are effectively switched 'off'. However, implementations of three-qubit gates have been shown and do require require simultaneously activated adjacent exchange couplings~\cite{Hendrickx2021, Madzik2022}. Moreover, quantum simulation of the rich variety of physical phenomena described by the Fermi-Hubbard model~\cite{Stafford1994, Hensgens2017,Barnes2019, Dehollain2020,Throckmorton2021, Wang2022,Knoerzer2022,Buterakos2023, Wang2023, NicoKatz2023} does require dense arrays of quantum dots with precise, and ideally orthogonal control not only of the on-site potentials but also of the nearest-neighbor exchange interactions or tunnel couplings. 

To appreciate why this is not straightforward, we note that once three or more spins are coupled together, the resulting energy spectrum and, hence, the oscillation frequencies are typically a combination of all the exchanges involved, hindering independent calibration of the exchange couplings. Furthermore, whereas the local electrochemical potentials vary linearly in the gate voltages, the tunnel coupling and exchange interaction depend exponentially on gate voltage. To realize cross-talk compensation in the face of this exponential dependence, references~\cite{Diepen2021, Qiao2020}  assumed that barrier-to-barrier cross-talk can be compensated by a linear combination of voltages in the $\emph{argument}$ of the exponential function, reducing cross-talk compensation between barriers to a linear problem nonetheless. In both cases, the virtualization methods required repeated measurements of either the tunnel coupling or exchange oscillations. Cross-talk was then extracted from exponential fits resulting in an indirect, laborious, and potentially error-prone measure. It is therefore desirable to obtain a measure of the cross-talk in a direct way similar to reservoir addition lines used to virtualize plunger gates. 

Here we demonstrate a way of characterizing barrier-to-barrier cross-talk to allow individual control of exchange interactions in a dense array of quantum dots.
The method consists of tracking the position of the singlet-triplet anti-crossing in the multi-dimensional voltage space spanned by the confining barrier gates. Such a feature can be induced by a suitable intrinsic spin-orbit interaction, local magnetic field gradients or differences in the g-tensors, and determines a point of constant exchange that is fast to measure and easy to identify. 
This is especially useful because it isolates the effect of the exchange interaction of interest and could also be adapted for automated optimizations~\cite{Mills2019a}. We empirically find how the gate architecture affects the cross-talk in the device. Finally, we apply this method to the tuning of four-spin Heisenberg chains in different configurations to test its validity and limitations.

\section{Device and energy diagram}
\begin{figure*}
    \centering
    \includegraphics[width = 0.95\textwidth]{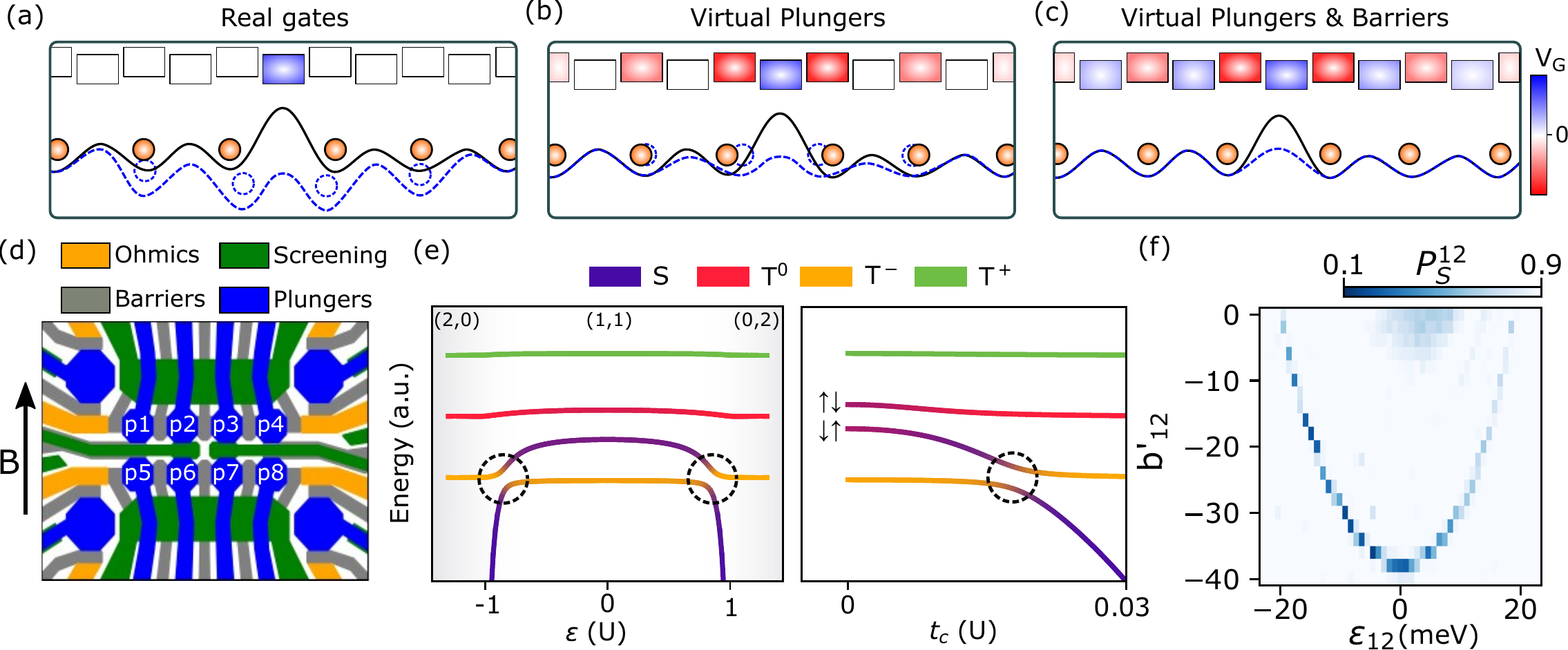}
    \caption{ (a) Schematics of the confinement potential for a chain of charges defined by the top gates in real voltage space (black solid line). A negative voltage pulse on the central barrier gate causes not only the middle tunnel barrier to be lowered but also shifts the electrochemical potentials of the nearby dots and the height of adjacent tunnel barriers (blue dashed line).
    (b) Commonly used virtual plunger gates work by applying a linear combination of gate voltages that keeps the electrochemical potentials of all other dots fixed. However, adjacent tunnel barrier heights are still affected and lateral shifts of charges are still present, although they might be slightly reduced.
    (c) If also the barriers are virtualized, a pulse on the middle barrier gate is compensated by suitable pulses on other barrier gates to keep the other tunnel barriers fixed and, ideally, counteract the lateral shifts of charges. In practice, however, only the combined effect of lateral shifts and tunnel barrier alterations can be compensated. A correct virtualization should allow orthogonal control of exchange interactions and enable a straightforward tuning of multi-spin chains.
    (d) Schematic of the 2$\times$4 dot array we use in this experiment. The dot plungers are labeled as $p_{i}$. Barriers $b_{ij}$ separate dots $i$ and $j$. The external magnetic field $B$ is applied in an in-plane direction marked by the arrow. 
    (e) Energy level diagram of a two-spin system in a double quantum dot as a function of detuning $\epsilon$ (left) and tunnel coupling $t_c$ (right). The dashed circles mark the spin-orbit induced avoided crossings. At $|\epsilon|>U$ the two charges occupy the same dot ((2,0) and (0,2) charge regions). For $|\epsilon|<U$ the charges are shared between the two adjacent dots and the energy splittings are determined by the respective Zeeman energies and the exchange interaction. The position of the avoided crossing can be influenced by $\epsilon$ and $t_c$. 
    (f) Measurement of the avoided crossing as a function of detuning and barrier voltage of $Q_{12}$, as described in the main text. The avoided crossing always occurs when $J=E_{T^-}$ constituting a constant-exchange feature which we are able to follow as a sharp reduction in singlet return probability $P_S^{12}$. At more positive values of $b_{12}'$, $S-T^0$ oscillations cause a reduced singlet return probability as well. As the barrier gets more negative, the exchange increases pushing the avoided crossing feature to smaller $\epsilon_{12}$. At $\epsilon_{12}=0$ all the exchange is induced by the barrier voltage $b_{12}'$.}
    \label{fig1}
\end{figure*}
The device consists of a 2x4 array of gate-defined quantum dots in a Ge/SiGe heterostructure~\cite{Lodari2021} (Fig.~\ref{fig1}d and further details can be found in the Appendix section~\ref{sec:device}).
Four sensors at the corners of the device enable fast charge sensing via radio-frequency (RF) tank circuits (the inductors are off-chip NbTiN coils while the capacitance stems from parasitic capacitances) bonded directly to one of the ohmic contacts of the respective sensor~\cite{Vigneau2023}. The potential landscape is tuned by means of DC voltages to form eight quantum dots, each containing a single hole, under the plunger gates $p_i$, with $i \in [1,8]$, except for dot 2 where, for practical reasons, we confine three holes. The interdot barrier gates $b_{ij}$ separate dots $i$ and $j$. Dots 1, 4, 5 and 8 have additional barriers to the reservoirs. Except for $b_{26}$ and $b_{37}$ all the barriers are deposited in the first gate layer allowing strong exchange tunability, unlike in previous experiments on 2x4 Ge/SiGe quantum dot arrays~\cite{Hsiao2024, Zhang2024} (see Appendix section~\ref{sec:J-profiles}). 
Importantly, since the charge carriers are holes, accumulation voltages are negative. A typical DC voltage configuration of the tuned up device is reported in Appendix Fig.~\ref{fig:DC_config}.
All the plunger gates and interdot barrier gates are connected also to an arbitrary waveform generator (AWG), via bias-tees and attenuated coaxial transmission lines, to allow fast pulsing away from the DC voltage configuration. In all measurements, the reported voltage amplitudes are the attenuated AWG amplitudes at the gates, without the DC component.

Throughout the experiments we work with virtualized plunger gates ($p'_{i})$ which are designed to vary the electrochemical potential of dot $i$ while keeping the electrochemical potential of all other dots fixed~\cite{Volk2019,Mills2019a}. The barrier gates are at first virtualized against the electrochemical potentials only and we denote them as $b'_{ij}$. This ensures that a pulse on a virtual barrier keeps the dot potentials unchanged (see Fig.~\ref{fig1}a,b). We further define a detuning axis $\epsilon_{ij} = ap'_i-bp'_j$ and an electrochemical potential axis $\mu_{ij}=cp'_i+dp'_j$, with $a,b,c,d$ experimentally determined coefficients (see Appendix section~\ref{sec:crosstalk matrix} for details on the transformations between real and virtual gate voltages). Every double dot is capable of hosting a singlet-triplet ($S-T$) qubit~\cite{Levy2002,Petta2005,Jirovec2021} which we label $Q_{ij}$ with $i$ and $j$ (we choose $i<j$) denoting the first and second dot in the pair, respectively. To operate $S-T$ qubits, precise control of the exchange interaction $J_{ij}$ is required (in the following discussion we omit the indices $ij$ and reintroduce them when necessary). In quantum dots systems, $J$ originates from the wave-function overlap of and the Coulomb repulsion between neighbouring spins~\cite{Loss1998,Qiao2020} and can be controlled by means of $\epsilon$ and tunnel coupling $t_c$ taking the form:
$$J(\epsilon, b') = \frac{4t^2_c(b')U}{(U^2-\epsilon^2)},$$
for $J\ll U$, where $U$ is the charging energy~\cite{Zhou2024}. The tunnel coupling is itself a function of the barrier voltage, and because of cross-talk, also of the voltage on neighbouring barriers. This crosstalk we will seek to compensate for in section~\ref{sec:Exch_characterization} to obtain orthogonal control of exchange interactions. We point out we will not attempt to compensate for the effect of plunger voltages on the exchange, as the interdot detuning is an explicit and desired control knob for the exchange strength, especially during readout and initialization. 

A typical energy diagram as a function of $\epsilon$ with finite $t_c$ of an $S-T$ qubit is depicted in the left panel of Fig.~\ref{fig1}e. Unless indicated otherwise, we operate every qubit at its symmetry point $\epsilon = 0$  where the exchange reduces to $J(b')=\frac{4t_c^2(b')}{U}$ and is, therefore, only controlled by the barrier voltage~\cite{Martins2016}. The energy diagram in this case is depicted in the right panel of Fig.~\ref{fig1}e.

Contrary to previous works~\cite{Hsiao2024}, here we do not measure $t_c$ and $U$, rather, we assume an empirical dependence of $J$ on the designated barrier: 
$$J(b') = J_0\mathrm{exp}(k(b'-b'_0)),$$
where $J_0= \SI{1}{\mega\hertz}$, $k$ represents the exponential lever arm of the barrier and $b'_0$ is an offset which depends on the particular DC voltage configuration. The DC configuration of the barriers is tuned in a way that ensures $b'_0$ to be relatively small such that the voltage pulses from the AWG are capable of inducing a considerable on-off ratio for each exchange (in general we find $|b'_0|< \SI{40}{\milli\volt}$, see Appendix section~\ref{sec:J-profiles}). 
Assuming zero residual exchange at the symmetry point the four eigenstates are the polarized triplets $\ket{T^-}= \ket{\downarrow\downarrow}$, $\ket{T_+} = \ket{\uparrow\uparrow}$ with energies $E_{T^\pm} = \pm \sum E_Z = \pm\sum g \mu_B B$, and the anti-parallel states $\ket{\uparrow\downarrow}$, $\ket{\downarrow\uparrow}$ with energies $E_{AP} = \pm \frac{\Delta E_Z}{2}=\pm \frac{\Delta g \mu_B B}{2}$. $g$ is the effective g-factor of each of the dots which we measure to take values between 0.3 and 0.45 in the in-plane magnetic field direction, consistent with previously reported values for holes in Ge~\cite{Hendrickx2021}. $\mu_B$ is the Bohr magneton and $B= \SI{10}{\milli\tesla}$ is the external magnetic field approximately applied in the in-plane direction. At large exchange $J$, the antiparallel states are no longer eigenstates of the Hamiltonian being replaced by $\ket{S} = \frac{\ket{\uparrow\downarrow}-\ket{\downarrow\uparrow}}{\sqrt{2}}$ and $\ket{T^0} = \frac{\ket{\uparrow\downarrow}+\ket{\downarrow\uparrow}}{\sqrt{2}}$. Finally, the spin-orbit spin-flip term $\Delta_{SO}$ couples $\ket{S}$ and $\ket{T^-}$ as highlighted by the dashed circles in Fig.~\ref{fig1}e~\cite{Jirovec2022, SaezMollejo2025}. At these avoided crossings, coherent $S-T^-$ oscillations can be induced~\cite{Mutter2021, Zhang2024}.

In the (2,0) charge configuration, the energy splitting between $\ket{S(2,0)}$ and $\ket{T(2,0)}$ well exceeds the thermal energy, enabling fast initialization in the singlet ground state.
The different energy scales of the system then allow us to initialize target eigenstates by appropriately choosing the ramp-time from the (2,0) to the (1,1) charge symmetry point (see Appendix section~\ref{sec:Init and read}). For read-out, we rely on Pauli spin blockade (PSB).
The rather small external magnetic field ensures that only the singlet state is unblocked in the PSB region, as opposed to parity readout where both anti-parallel spin states are typically unblocked~\cite{Niegemann2022}. This enables the discrimination of $\ket{S}$ from the triplets by monitoring the charge sensor in single-shot readout.

In Fig.~\ref{fig1}f we experimentally map out the $ST^-$ avoided crossing as a function of $\epsilon_{12}$ and $b'_{12}$ by initializing $\ket{S}$ in $Q_{12}$ and recording the probability to retrieve $\ket{S}$. In between, we rapidly pulse to (1,1), let the system evolve for $\SI{50}{\nano\second}$, corresponding approximately to flipping the initial $\ket{S}$ to $\ket{T^-}$ at the avoided crossing, and pulse back to (2,0) for readout. We can identify the position of the $ST^-$ avoided crossing by a sharp reduction in $P_S$. Since at these positions $J(\epsilon, b')= E_{T^-}$, the avoided crossing constitutes a constant-exchange feature. The symmetric U-shape is a confirmation that the virtualization of $b'_{12}$ against the plunger gates is accurate as we would otherwise find a skewed shape~\cite{Zhang2024} (see Appendix Fig.~\ref{fig:skewed_spin_cup}). For more positive barrier voltages we also observe a reduction of $P_S$ which can be attributed to $S-T^0$ oscillations at low $J$. Importantly, around the avoided crossing, there is no other feature present which allows for a precise identification of its location. 

\section{Exchange cross-talk}
\begin{figure}
    \centering
    \includegraphics[width=0.5\textwidth]{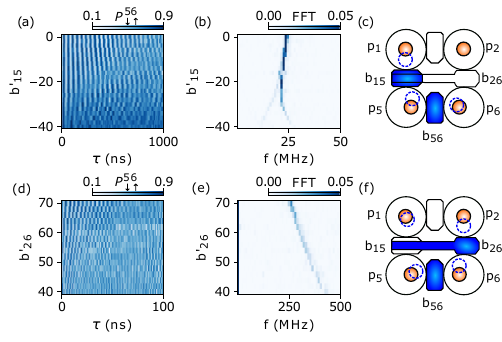}
    \caption{(a) Exchange oscillations in $Q_{56}$ as a function of $b'_{15}$ and dwell time $\tau$, as explained in the main text. 
    (b) FFT of (a). We clearly see that the main frequency is reduced and for more negative values of $b'_{15}$ a second oscillation frequency appears. The frequency reduction is a sign of cross-talk, the appearance of a second frequency is due to a finite $J_{15}$.  
    (c) Sketch of the experiment in (a). The orange circles depict the approximate charge positions when only $b'_{56}$ induces exchange. The blue dashed circles represent the shifted dot positions as we open $b'_{15}$.
    (d) Exchange oscillations in $Q_{56}$ as a function of $b'_{26}$ and dwell time $\tau$. 
    (e) FFT of (d). We see a change in frequency of about $60\%$ over only $\SI{30}{\milli\volt}$, indicating strong cross-talk.
    (f) Sketch of the experiment in (d) similar to (c). The fan-out of $b_{26}$ leads to a much larger cross-talk than for $b_{15}$ and may affect the position of all the nearby charges.}
    \label{fig:Typ_exch}
\end{figure}

Before discussing how to correct for cross-talk, we want to elucidate how exchange interactions are affected by nearby barrier voltages. As an example, consider the left side of the device. Fig.~\ref{fig:Typ_exch}a shows exchange oscillations in $Q_{56}$ as a function of $b'_{15}$, with $b'_{56}$ kept at a constant, negative value (see sketch in Fig.~\ref{fig:Typ_exch}c). We initialize $Q_{56}$ in $\ket{\downarrow\uparrow}$ and $Q_{12}$ in $\ket{\downarrow\downarrow}$ and record the final state probability $P_{\downarrow\uparrow}^{56}$ as a function of dwell time $\tau$. At first, $b'_{56}$ is the only gate inducing exchange between the spins in dot 5 and 6 leading the oscillation seen in the top part of Fig.~\ref{fig:Typ_exch}a. As we pulse $b'_{15}$ more negative, we first see the frequency of the oscillations reduce, a clear example of cross-talk. Furthermore, we also see another frequency appearing below $b'_{15} = -\SI{20}{\milli\volt}$ (see also the FFT in Fig.~\ref{fig:Typ_exch}b). This is an indication that $b'_{15}$ now induces exchange between dots 5 and 1. In such a situation it is not clear which of the measured frequency shifts can be attributed to the activation of $J_{15}$ or to cross-talk on $J_{56}$. Without further modeling, it is, therefore, not possible to quantify the effect of $b'_{15}$ on $J_{56}$ in this voltage range.
We can only reliably extract the cross-talk for values $b'_{15}>-\SI{20}{\milli\volt}$, e.g. before we induce any measurable exchange $J_{15}$.

A similar experiment is plotted in Fig. \ref{fig:Typ_exch}d-f. However, in this case we pulse on $b'_{26}$ ($b'_{56}$ is more negative than in Fig.~\ref{fig:Typ_exch}a leading to much faster oscillations). While here we do not see another frequency appear because we do not apply a large enough pulse on $b'_{26}$, we observe a change in frequency of about $60 \%$ in a range of only $\SI{30}{\milli\volt}$, indicating very strong cross-talk. This might be due to the fan-out of $b_{26}$ as highlighted in the sketch in Fig.~\ref{fig:Typ_exch}f. While this measurements reveals the cross-talk from $b'_{26}$ onto $J_{56}$, it doesn't tell us how to compensate for it. In the next section we show how exchange cross-talk can be extracted directly and compensated.

\section{Characterization of exchange virtualization parameters}
\label{sec:Exch_characterization}
\begin{figure*}
    \centering
    \includegraphics[width=\textwidth]{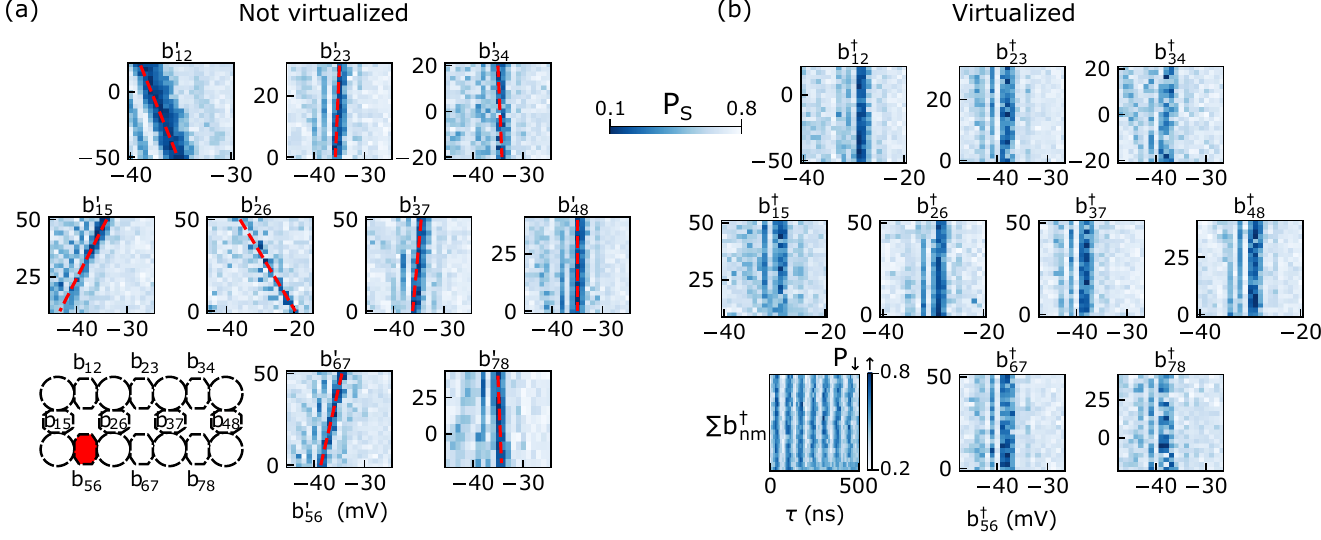}
    \caption{(a) $ST^-$ avoided crossing of $Q_{56}$ as a function of $\mathrm{b'}_{56}$ on the horizontal axis and all the other barriers on the respective vertical axis. The plots are ordered to reflect the geometric location of the stepped gate. The position of the avoided crossing is reflected by a sharp decrease of the singlet return probability (see main text). From the fitted red dashed lines we extract the cross-talk elements $\alpha_{56}^{mn}$. The fact that we can fit all cross-talk features with a linear function confirms the assumption of linear barrier cross-talk.
    (b) $ST^-$ avoided crossing of $Q_{56}$ as a function of $\mathrm{b}^\dagger_{56}$ on the horizontal axis and all the other virtual barriers on their respective vertical axis. After the virtualization process the $ST^-$ avoided crossing position is only controlled by $b^\dagger_{56}$ as intended. To further verify that the exchange remains stable we plot exchange oscillations of $Q_{56}$ in the bottom left. We vary all the virtual barriers except $\mathrm{b}^\dagger_{56}$ together in the same range as in the individual plots. As desired, the exchange oscillations do not change in the ranges considered here. We repeat the same procedure on the other barrier gates (see Appendix section \ref{sec:Cross-talk other barriers}) and successively fill in the values for $\alpha_{ij}^{mn}$. 
    }
    \label{fig:Fig2}
\end{figure*}

We now show how we can leverage the $S-T^-$ avoided crossing to directly extract the cross-talk matrix element for barrier to barrier cross-talk compensation. At the avoided crossing, with $\epsilon_{ij}= 0$, all the exchange is induced by the virtual barrier gate voltage $b'_{ij}$ satisfying $|J(b'_{ij})|=|E_{T_{ij}^-}|$ and is to first order insensitive to small variations of $\epsilon_{ij}$. Moreover, any unintentional variation in $\epsilon_{ij}$ will only increase $J_{ij}$, but never decrease it as in the case of lateral shifts of the dot positions. To compensate exchange cross-talk we introduce a second layer of virtualization and define new virtual barrier voltages as $b^\dagger_{ij}$. As in previous works\cite{Qiao2020,Diepen2021} we assume a linear barrier cross-talk and an exponential dependence of $J$ on the new virtual barrier: $J_{ij} = J_{0}\exp(k (b^\dagger_{ij}-b^\dagger_{0ij}))$, $b^\dagger_{ij} = \sum_{mn} \alpha_{ij}^{mn} b'_{mn}$, where $nm$ are all the tuples corresponding to adjacent spins and $\alpha_{ij}^{mn}= \frac{\delta J_{ij}}{\delta b'_{mn}}/ \frac{\delta J_{ij}}{\delta b'_{ij}}$. The term $b^\dagger_{0ij}$ is an offset in $b^\dagger_{ij}$ that we need to quantify only when calibrating the dependence of $J$ on $b^\dagger_{ij}$. For cross-talk compensation, we therefore need to determine all the values $\alpha_{ij}^{mn}$, where by definition $\alpha_{ij}^{ij}=1$. Since the avoided crossing constitutes a constant-exchange feature, we can track its position as a function of $b'_{mn}$ and $b'_{ij}$ and extract a slope returning $-\alpha_{ij}^{mn}$ directly, without the need to extract $\frac{\delta J_{ij}}{\delta b'_{mn}}$ with exponential fits through a series of datapoints. This is the main advantage of the method presented here.

Fig. \ref{fig:Fig2}a shows measurements of the avoided crossing of $Q_{56}$ as a function of $b'_{56}$ and all other $b'_{mn}$. 
The position of the avoided crossing is reflected by a sharp reduction in $P_S$. In all plots we can follow this stand-out feature with Gaussian fits and extract the red dashed lines (see Appendix). The linear slopes confirm the assumption that barrier gate cross-talk is linear, at least in this regime. 
The value of the slope $\frac{\delta b'_{56}}{\delta b'_{mn}}$ directly returns the cross-talk element $\alpha_{56}^{mn}$.

By plugging $\alpha_{56}^{mn}$ in the correction matrix and repeating the measurement of Fig. \ref{fig:Fig2}a as a function of $b^\dagger_{56}$ and $b^\dagger_{mn}$ (Fig. \ref{fig:Fig2}b), we now observe completely vertical constant-exchange features, controlled exclusively by $b^\dagger_{56}$, as intended. The bottom left panel in Fig. \ref{fig:Fig2}b further confirms that the cross-talk is compensated as we record exchange oscillations of $Q_{56}$ as a function of all the $b^\dagger_{mn}$ except $b^\dagger_{56}$ and observe no change in frequency for the voltage ranges considered here. Note that the voltage range of $b^\dagger_{12}$ we scan here corresponds to an on-off ratio of $J_{12}$ of $>100$. Similar ratios apply to $J_{23}$, $J_{34}$, $J_{48}$, and $J_{78}$ and their respective virtual barrier gates. This shows that for non-adjacent exchange interactions, the gate cross-talk can be efficiently compensated over at least two orders of magnitude. The virtual barriers $b^\dagger_{15}$, $b^\dagger_{26}$, and $b^\dagger_{67}$ are scanned over a range chosen as to not induce any exchange, since this would alter the position of the avoided crossing even without cross-talk (recall Fig.~\ref{fig:Typ_exch}). We will show in section \ref{sec:validation} how to test whether the virtualization remains effective also when adjacent exchanges are turned on.

We find correction factors for every gate where we were able to induce exchange, always taking care that nearest-neighbor gates do not induce any exchange. For practical reasons we did not characterize virtualization of $b^\dagger_{26}$, $b^\dagger_{37}$ and $b^\dagger_{67}$ but only their effect on other gates. In fact, the fan-out of $b_{26}$ and $b_{37}$ affects the charge sensors and result in a loss of the read-out signal, while a too negative voltage in $b_{67}$ accumulates spurious dots that couple to the spins in dots 5 and 6. These problems will be addressed in future device generations.

\begin{figure*}
    \includegraphics[width = 0.9\textwidth]{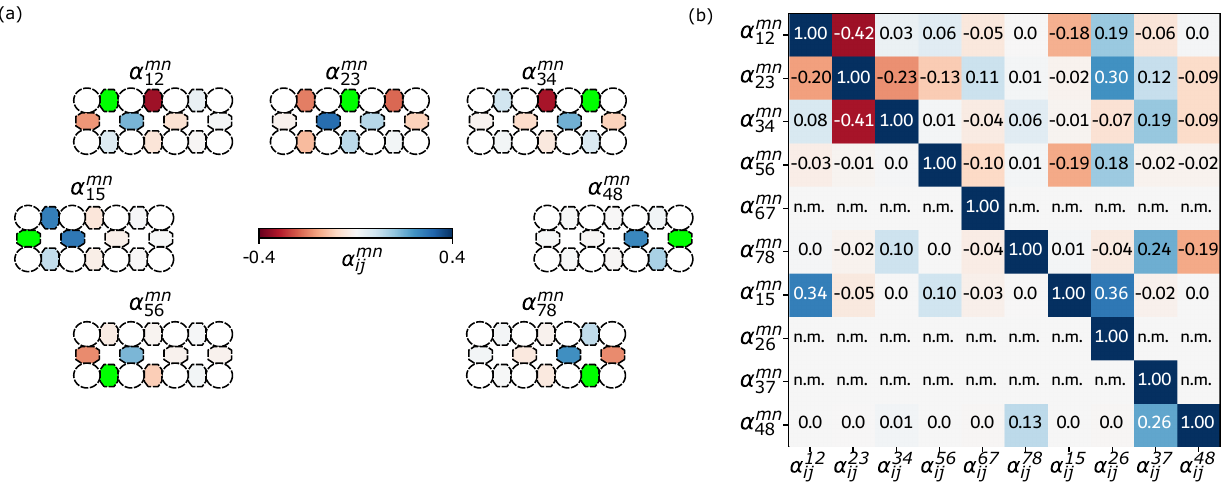}
    \caption{(a) Summary of the exchange cross-talk elements $\alpha_{ij}^{mn}$ extracted for the barriers highlighted in bright green. We clearly see a reduction of cross-talk with distance as we expect for capacitive cross-talk. Positive (negative) cross-talk elements imply that an adjacent barrier gate enhances (reduces) a given exchange coupling. 
    (b) Virtual gate matrix for the barrier gates  showing the $\alpha_{ij}^{mn}$, summarizing the results from (a). Except for $\alpha_{26}^{mn}$, $\alpha_{37}^{mn}$, and $\alpha_{67}^{mn}$, all the elements $\alpha_{ij}^{mn}$ are measured. In the experiment the elements labeled 'n.m.' are replaced by 0. To be clear, this matrix reports how much each $b'_{ij}$ affects the various $b^\dagger_{mn}$. The barrier gate voltages $b'_{ij}$ needed to orthogonally control the respective exchange interactions via $b^\dagger_{mn}$ are obtained from the inverse of this matrix. }
    \label{fig:AllVirtAndMatrix}
\end{figure*}

Fig.~\ref{fig:AllVirtAndMatrix}a color codes the extracted cross-talk elements and shows their geometrical distribution with respect to the gate which is being virtualized (highlighted by  the bright green color). 
We clearly observe a decay of $|\alpha_{ij}^{mn}|$ with distance as expected. We note that cross-talk between barriers that are side by side along the legs of the ladder is negative, similar to previous works~\cite{Diepen2018, Qiao2020}. In contrast, cross-talk between barriers oriented orthogonal to each other can also be positive, which is an indication that lateral shifts of the dot positions are less important. Finally, many $\alpha_{ij}^{26}$ and $\alpha_{ij}^{37}$ are relatively large, sometimes exceeding 30$\%$ (see Fig.~\ref{fig:AllVirtAndMatrix}b), which suggests that the fan-out of their respective gates, $b_{26}$ and $b_{37}$, can induce considerable cross-talk (see Fig.~\ref{fig1}d and Appendix Fig.~\ref{fig:Design}). This fact should be taken into account when designing future devices.

In the cross-talk extraction we have not considered $g$-factor modulation due to detuning and barrier gate voltages~\cite{Mauro2024}. Moreover, we note that occasionally we have to slightly correct the initially extracted virtualization matrix element (Fig. \ref{fig:AllVirtAndMatrix}b) to obtain an accurate cross-talk compensation.

Finally, we note that this method is suitable when exchange values are of the same order as the Zeeman energy. For $|J| \ll |E_{T^-}|$, which is the case for high external magnetic field operation, there is no avoided crossing to follow. To circumvent this, one could determine the cross-talk elements at low external field and then perform the desired experiments at higher field. Alternatively, it might be possible to use microwave driving to track the dependence of the spin resonance frequency. Indeed, the resonance frequency of a spin is modified by the exchange interaction, so it would be possible to fix the applied microwave tone and scan the desired barrier against all other barriers and extract $\alpha_{ij}^{mn}$ in the same way we did here.

In the following we explore whether $\alpha_{ij}^{mn}$ is stable even when nearest-neighbor exchanges are turned on. This is a crucial question as it would allow to extend spin chains indefinitely after locally mitigating cross-talk.

\section{Validation of the virtualization in extended spin-chains}
\label{sec:validation}
In order to test if the virtualization parameters are still valid in a regime where nearest-neighbor exchanges are turned on, we proceed to couple four spins into a chain and observe the resulting oscillation dynamics. As shown in Fig. \ref{fig:Typ_exch}a,b, activated nearest-neighbor interactions result in a highly non-trivial time evolution with multiple visible oscillation frequencies. However, preparing special states and evolving them at special voltage points results in a single visible oscillation frequency which we use here to validate the virtualization. For a chain of coupled spins, the Hamiltonian describing the system can be approximated as
\begin{equation}
    H = \sum_{i}g_i\mu_B B S_{z,i}+\sum_{i}\Delta_{SO,i} S_{x,i}+\sum_{<i,j>} J_{ij}\left(\bm{S_iS_j}-\frac{1}{4}\right)
\label{eq:4spinHam}
\end{equation}
where $\mathbf{S}=(S_{x}, S_{y}, S_{z}) = \hbar/2 (\sigma_x, \sigma_y, \sigma_z)$ is the vector of spin operators with the Pauli matrices $\sigma_{x,y,z}$ for each spin and the indeces $i,j$ run over nearest neighbors, and we set $\hbar = 1$. The additional term $\Delta_{SO}$ stems from intrinsic spin-orbit interaction and the anisotropic g-tensors~\cite{Zhang2024}.
Ignoring the spin-orbit term and considering four adjacent spins, Eq. \ref{eq:4spinHam} can be conveniently written in the reduced basis $\bigl\{ \ket{S_{ij}S_{kl}}, \ket{S_{ij}T^-_{kl}}, \ket{T^-_{ij}S_{kl}}, \ket{T^-_{ij}T^-_{kl}}\bigr\}$ as $H = H_Q=$ 
\begin{equation*}
    \left( 
    \begin{matrix}
    -J_{ij}-J_{kl} & 0 & 0 & 0 \\
    0 & -J_{ij}-\overline{E}_{z,kl} & -\frac{J_{jk}}{2} & 0 \\
    0 & -\frac{J_{jk}}{2} & -\overline{E}_{z,ij}-J_{kl} & 0 \\
    0 & 0 & 0 & -\overline{E}_{z,ijkl}+\frac{J_{jk}}{2}
    \end{matrix}\right)
    \label{HQ}
\end{equation*}

We want to draw attention to the matrix elements that involve $\ket{S_{ij}T^-_{kl}}$ and $\ket{T^-_{ij}S_{kl}}$. When $|J_{ij}-\overline{E}_{z,ij}|=|J_{kl}-\overline{E}_{z,kl}|$, the diagonal elements are equal and the resulting degeneracy is lifted by the off-diagonal elements $-J_{jk}/2$. We call this the resonant $ST^-$ condition. Initializing one of the two states $\ket{S_{ij}T^-_{kl}}$ or $\ket{T^-_{ij}S_{kl}}$, and letting the system evolve at this special point, results in $\ket{ST^-}\leftrightarrow \ket{T^-S}$ oscillations with a frequency  $hf = J_{jk}$. This convenient feature was exploited in \cite{Zhang2024} to implement a two-qubit gate for singlet-triplet qubits and in \cite{Wang2023} to estimate the exchange interaction in a four-qubit plaquette. Similar arguments hold for the $\bigl\{ \ket{S_{ij}S_{kl}}, \ket{S_{ij}T^0_{kl}}, \ket{T^0_{ij}S_{kl}}, \ket{T^0_{ij}T^0_{kl}}\bigr\}$ subspace where the resonant condition appears when $\sqrt{J_{ij}^2+\Delta{E}^2_{z,ij}}=\sqrt{J_{kl}^2+\Delta{E}^2_{z,kl}}$. We utilize these resonant conditions to verify whether the virtualization obtained with nearest-neighbour couplings switched off is still valid when nearest-neighbor interactions are turned on. This test relies on the fact that the position of the resonant condition depends on both $J_{ij}(b^\dagger_{ij})$ and $J_{kl}(b^\dagger_{kl})$ (as well as the Zeeman energies), while the oscillation frequency depends on $J_{jk}(b^\dagger_{jk})$. Hence, we can test whether we find the resonant condition at the correct location and whether the oscillation frequency matches our expectations (more details can be found in Appendix section~\ref{sec:ST resonant cond}).

We study three implementations of a four-spin chain with nearest-neighbor couplings: chain 3-4-8-7, chain 2-1-5-6, and chain 1-2-3-4. In the first two cases, the chains are curved around the right and left edges of the device, respectively, while in the last case the dots forming the chain are assembled linearly. The latter situation was previously studied in GaAs devices in \cite{Qiao2020, Diepen2021}. 

\begin{figure*}
    \includegraphics[width = 0.85\textwidth]{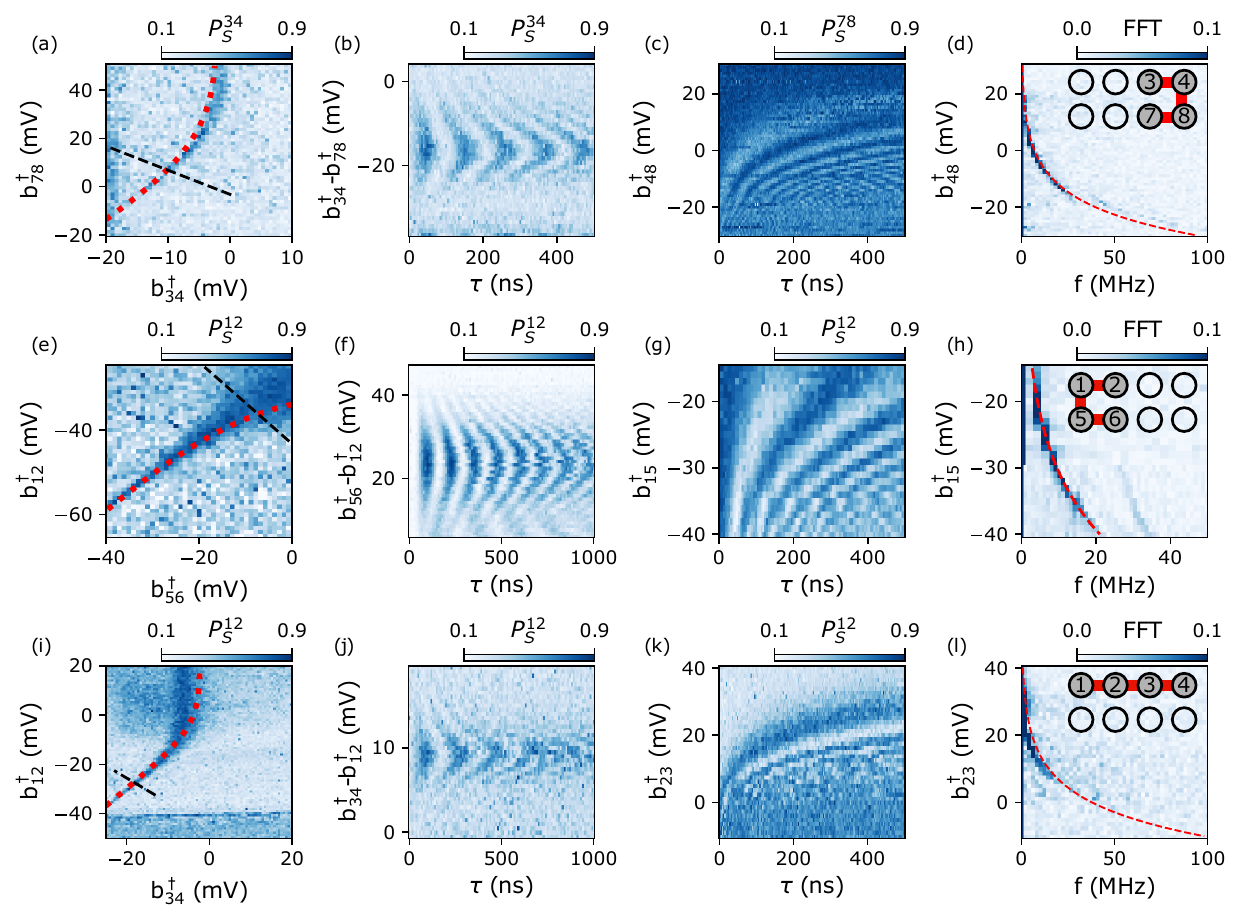}
    \caption{
    (a) Resonant $\ket{S_{78}T^-_{34}}\leftrightarrow \ket{T^-_{78}S_{34}}$ condition as a function of $b^\dagger_{34}$ and $b^\dagger_{78}$ with an exchange $J_{48}\approx \SI{2}{\mega\hertz}$ induced by $b^\dagger_{48}$. We record the probability of measuring $\ket{S_{34}}$ after initializing $\ket{S_{78}T^-_{34}}$ and letting the system evolve for $\tau = \SI{380}{\nano\second}$ corresponding to a near perfect inversion of population at the resonant condition marked by an increase in $P_{S}^{34}$. The red dots mark the theoretical resonant condition, based on the Zeeman energies and individual exchange dependencies, which agrees well with the data.
    (b) Resonant $\ket{S_{78}T^-_{34}}\leftrightarrow \ket{T^-_{78}S_{34}}$ oscillations as a function of dwell time $\tau$ and $b^\dagger_{34}-b^\dagger_{78}$. The barriers are scanned along the dashed line in (a). The maximum oscillation amplitude corresponds to the resonant condition and the frequency is $hf = J_{48}$.
    (c) Resonant $\ket{S_{78}T^-_{34}}\leftrightarrow \ket{T^-_{78}S_{34}}$ oscillations as a function of dwell time $\tau$ and $b^\dagger_{48}$ at the resonant condition. 
    (d) FFT of (c). The red dashed line is the exchange dependence $J_{48}(b^\dagger_{48})$ extracted from the isolated $Q_{48}$ oscillations which matches well with the observed FFT peak. The latter yields $J_{48}(b^\dagger_{48})$ with the two neighbouring exchanges activated. The inset shows a sketch of the dots and interactions involved in the experiments (a)-(d).
    (e)-(h) Similar to (a)-(d) but for chain 2-1-5-6. In this case $b^\dagger_{15}$ was not virtualized due to the rather small cross-talk elements. (g) and (h) show data pertaining to the resonant $\ket{S_{12}T^0_{34}}\leftrightarrow \ket{T^0_{12}S_{34}}$ condition. The red dashed line in (h) is a fit and allows us to extract $J_{15}(b^\dagger_{15})$ even without complete virtualization.
    (i)-(l) Similar to (a)-(d) but for chain 1-2-3-4. While we find good agreement with the predicted resonant condition in (i), the oscillation frequency in (j) and (k) is different from the expected value extracted from the isolated $Q_{23}$ oscillations. We do find good agreement with the data if we correct the value $b^\dagger_{0,23}$ by $-\SI{6}{\milli\volt}$. This suggests that some residual, possibly non-linear cross-talk remains, which will require more sophisticated mitigation strategies to account for.
    }
    \label{fig:ResonantConditions}
\end{figure*}

We first consider the chain 3-4-8-7. After having extracted the cross-talk coefficients $\alpha_{34}^{mn}$, $\alpha_{48}^{mn}$, and $\alpha_{78}^{mn}$ we measure exchange oscillations for $Q_{34}$, $Q_{48}$, and $Q_{78}$ as a function of $b^\dagger_{34}$, $b^\dagger_{48}$, and $b^\dagger_{78}$, respectively, to extract the dependence $J_{ij}(b^\dagger_{ij})$. For these measurements, the other exchange interactions are turned off such that the oscillation frequency only depends on the exchange interaction of interest.
We then fit the oscillation frequency to 
\begin{equation}
    f_{ij}(b^\dagger_{ij}) = \sqrt{\left(J_{0}\exp{k_{ij}(b^\dagger_{ij}-b^\dagger_{0,ij})}\right)^2+\Delta E_{Z,ij}^2}
    \label{eq:Exchange_frequency}
\end{equation}
with $k_{ij}$, $b^\dagger_{0,ij}$, and $\Delta E_{Z,ij}$ as free parameters (see Appendix Fig. \ref{fig:exch_profiles}). With the knowledge of these exchange dependencies as well as the Zeeman energies, we are able to predict at which voltage points $(b^\dagger_{34}, b^\dagger_{78})$ the resonant conditions should appear.
Fig. \ref{fig:ResonantConditions}a shows the resonant $ST^-$ condition as a function of $(b^\dagger_{34},b^\dagger_{78})$ with the exchange in between set to $J_{48}\approx \SI{2}{\mega\hertz}$ through $b^\dagger_{48}$. We initialize $\ket{S_{78}T^-_{34}}$ and let the system evolve for $\tau = \SI{380}{\nano\second}$ at each voltage point which ensures an approximate population inversion to $\ket{T^-_{78}S_{34}}$ at the resonant condition as long as $J_{48}$ remains unaffected by $b^\dagger_{34,78}$.
We identify the resonant condition as a sharp change in return probability. We do not record the joint probability of measuring $\ket{S_{34}T^-_{78}}$ but rather choose to measure only $P_{S}^{34}$. By operating in regimes where leakage outside the $\ket{ST^-}, \ket{T^-S}$ subspace is suppressed, we still recover the desired information. The red dotted line is the predicted location of the resonant condition based on the extracted exchange dependencies and the Zeeman energies, which agrees well with the data. Numerical simulations of $P_{S}^{34}$ resulting from the full system dynamics also match the experimental data very well (Appendix Fig. \ref{fig:3478_Sims}a). This suggests that $J_{34}(b^\dagger_{34})$ and $J_{78}(b^\dagger_{78})$ are still well virtualized even when $J_{48}$ is activated.

To extract the value of $J_{48}$ at a given voltage point, we can record the dynamics at the resonant condition. To do this, we choose any point along the resonant condition, away from any leakage features, and sweep $b^\dagger_{34}$ ($b^\dagger_{78}$) by $\pm \SI{5}{\milli\volt}$ ($\mp \SI{5}{\milli\volt}$), here resulting in the black dashed line in Fig. \ref{fig:ResonantConditions}a, as a function of dwell time. In \ref{fig:ResonantConditions}b we clearly see a chevron pattern with a maximum in amplitude, corresponding to the resonant condition, oscillating at a frequency given by $J_{48}$. This frequency agrees well with the corresponding frequency seen in numerical simulations (Appendix Fig. \ref{fig:3478_Sims} b), providing evidence that also $b^\dagger_{48}$ is properly virtualized.

Finally, in Fig. \ref{fig:ResonantConditions}c we report coherent $\ket{S_{78}T^-_{34}}\leftrightarrow\ket{T^-_{78}S_{34}}$ oscillations as a function of $b^\dagger_{48}$ at the resonant condition, while recording $P_S^{78}$ this time. Fig. \ref{fig:ResonantConditions}d is the FFT of (c) and the red dashed line is the exchange dependence $J_{48}(b^\dagger_{48})$ we extracted from the isolated $Q_{48}$ measurements. Since also here we find good agreement, we conclude that, in this case, the virtualization of all three barrier gates involved was successful.

We repeat the same procedure for chain 2-1-5-6 (Fig.~\ref{fig:ResonantConditions}e-h). Here we show data that was taken during a different cool-down and where $b^\dagger_{15}$ was not virtualized ($b^\dagger_{12}$ and $b^\dagger_{56}$ were virtualized). However, the cross-talk onto $b^\dagger_{15}$ turned out to be relatively weak such $J_{15}$ was only mildly affected by $b^\dagger_{12}$ and $b^\dagger_{56}$. Moreover, Fig.~\ref{fig:ResonantConditions}g-h pertain to the resonant $\ket{ST^0}$ condition. The red dashed line in Fig. \ref{fig:ResonantConditions}h is a fit to the data that allows us to extract $J_{15}(b^\dagger_{15})$ for this particular voltage configuration. In addition, in Appendix Fig. \ref{fig:1256_resCond_vs_Vvb15} we report further resonant condition plots similar to Fig.~\ref{fig:ResonantConditions}f for different values of $b^\dagger_{15}$ and observe that the position of the resonant condition does not change. We therefore conclude that the virtualization of $b^\dagger_{12}$ and $b^\dagger_{56}$ is correct also in this four-spin chain, and we have full knowledge of the Hamiltonian parameters in this configuration. 

For the ultimate test, we consider the same plots as for 3-4-8-7 also for the chain 1-2-3-4 involving $b_{23}$, which displays the most severe cross-talk elements $\alpha_{23}^{mn}$. (Fig.~\ref{fig:ResonantConditions}i-l). Although we find a good theoretical agreement with the data in Fig. \ref{fig:ResonantConditions}i, to match the oscillation frequencies in \ref{fig:ResonantConditions}j,k, we needed to adjust the value of $b_{0,23}^\dagger$ by $-\SI{6}{\milli\volt}$ compared to the values extracted via Eq.~\ref{eq:Exchange_frequency} from the isolated oscillations $Q_{23}$. This suggests that, while we are able to correct for most of the cross-talk, there might be some non-linear effects which we did not take into account. We have identified a possible cause for this, which was also extensively discussed in ref.~\cite{Rao2025}. In fact, we observe a nonlinear cross-talk between $b'_{23}$ and $p'_3$ which we show in Appendix Fig.~\ref{fig:p3vsvb23}. This can lead to a miscalibrated cross-talk and the observed discrepancy between the isolated $Q_{23}$ and the coupled four-spin chain measurement. To correct for such effects, we would require a more sophisticated, non-linear cross-talk compensation scheme which is outside the scope of this work. 

In general, we find that we are able to compensate most of the cross-talk and infer the Hamiltonian parameters also in the coupled four-spin chains. 
All the data is supported by numerical simulations which we report in Appendix section~\ref{sec:Sims}.

\section{Discussion}
In this work, we have proposed and demonstrated a way of directly extracting barrier-to-barrier cross-talk by tracking the constant-exchange feature given by the $ST^-$ avoided crossing in Ge. While the effects generally vary from gate to gate, we were able to observe a few trends. In fact, electrodes featuring a fan-out in proximity to other gates typically display large cross-talk to other exchanges. For two-dimensional quantum dot arrays, these lateral fan-outs can be avoided through a vertical fan-out with vias~\cite{Ha2021, Weinstein2023, George2024}. Perfecting the gate layout and oxide composition can also contribute to reduced cross-talk as demonstrated recently~\cite{Madzik2025}.
On the other hand, barrier gates oriented perpendicularly to each other and with no proximal fan-out, typically show much less cross-talk which seems easier to mitigate. This suggests that, for 1D chains, a zigzag alignment of quantum dots could be favorable over a strictly linear placement.

An important verification tool in our work is the construction of four-spin chains and their time evolution at the resonant $ST^-$ or $ST^0$ condition. It allowed us to confirm that the cross-talk extracted in the isolated two-qubit regime also mostly carries over to the regime of coupled four-spin chains. This important observation might enable the construction of longer spin-chains with only local cross-talk calibration. 

Finally, we want to emphasize also some of the limitations of this method. If we want to utilize the $ST^-$ avoided crossing, we can only calibrate the cross-talk reliably in the isolated two-spin regime and only for $J\approx E_{T^-}$. Other methods for direct cross-talk extraction can be utilized and also rely on following a constant-exchange feature, but are still limited to the isolated two-spin regime. For example, to extract the cross-talk elements $\alpha_{23}^{mn}$ we made use of exchange oscillations at a fixed time evolution (see Appendix Fig. \ref{fig:23virt}) as we did not calibrate $\ket{S}$ initialization and direct PSB readout due to the distance from the sensing dots. This method is more versatile as it doesn't require spin-orbit interactions, making it suitable also for GaAs or silicon quantum dot devices. Nonetheless, we found that the $ST^-$ avoided crossing is a feature that is typically robust and easy to measure without knowledge of the particular Hamiltonian parameters like the exchange dependence on any barrier nor any lever arms. This information was crucial in earlier works on virtualization \cite{Diepen2018, Qiao2020}. In practice the method presented here is, therefore, preferable as it requires less overall knowledge of the device. These considerations make this method also a good candidate for automated calibrations~\cite{Diepen2018,Mills2019a} which can be easily integrated into existing routines~\cite{Rao2025}.

With our findings, we have shed further light on the intricate cross-talk behavior in multi-layered spin qubit devices. We also demonstrated that, despite the density of electrodes, linear cross-talk can be managed and corrected for. Generally, the designing of large spin-qubit arrays leaves room for improvement, for which this work provides valuable guidance. Furthermore, our results open the possibility of observing multi-spin physics in longer chains and two-dimensional geometries, with detailed knowledge of the underlying Hamiltonian.

\section*{Acknowledgements}
We thank M. Rimbach-Russ, T.-K. Hsiao, V. John, F. Borsoi, C.-A. Wang and other members of the Vandersypen, Veldhorst, Scappucci, Rimbach-Russ and Bosco groups for stimulating discussions. We acknowledge technical support by O. Benningshof, J. D. Mensingh, T. Orton, R. Schouten, R. Vermeulen, R. Birnholtz, E. Van der Wiel, B. Otto and D. Brinkman. This work was funded by an Advanced Grant from the European Research Council (ERC) under the European Union’s Horizon 2020 research (882848). We acknowledge support by the European Union through an ERC
Starting Grant QUIST (850641). We acknowledge support by the Army Research Office (ARO) under grant number W911NF2310110. The views and conclusions contained in this document are those of the authors and should not be interpreted as representing the official policies, either expressed or implied, of the ARO
or the US Government. The US Government is authorized
to reproduce and distribute reprints for government purposes
notwithstanding any copyright notation herein.

\section*{Data availability}
The raw measurement data and the analysis supporting the
findings of this work are available on a 4TU.research repository~\cite{Datasets}.


\section*{Appendix}

\setcounter{page}{1}
\setcounter{section}{0}
\makeatletter
\renewcommand{\theequation}{\arabic{equation}}
\renewcommand{\thefigure}{\arabic{figure}}
\renewcommand{\thesection}{\arabic{section}}
\renewcommand{\thesubsection}{\arabic{subsection}}
\subsection{Device fabrication and experimental setup}
\label{sec:device}
\begin{figure*}[h]
    \centering
    \includegraphics[width=0.85\linewidth]{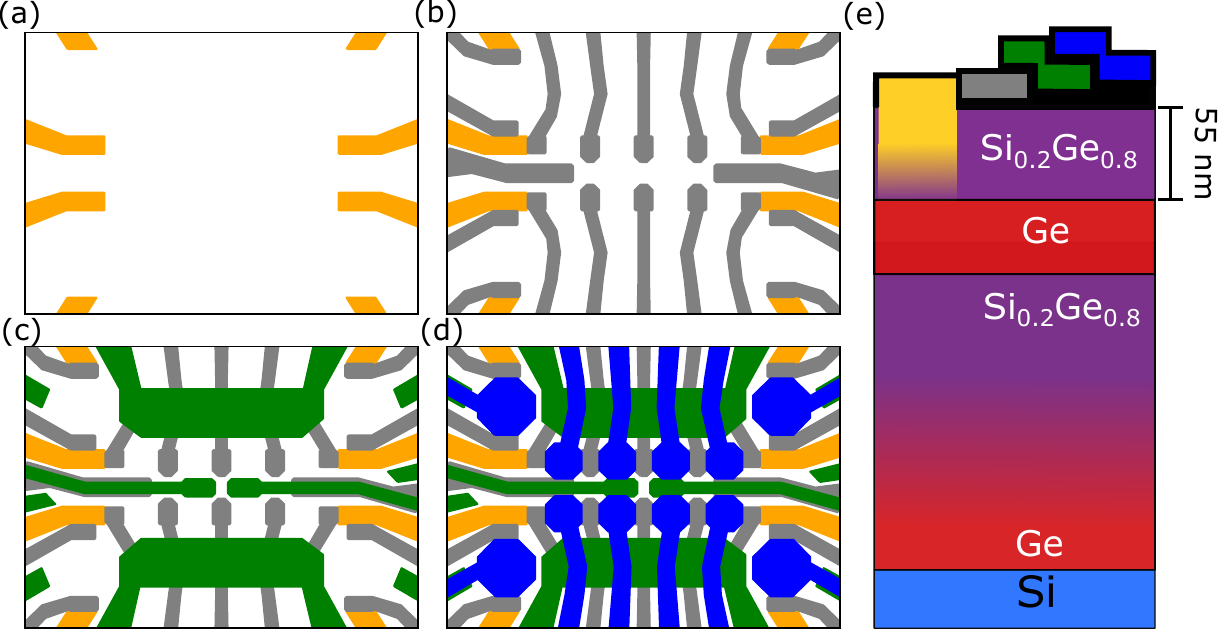}
    \caption{Device design and heterostructure. (a) In the first metal layer we deposit the Ohmic contacts. (b) After annealing while depositing the gate oxide, we proceed with the first barrier gate layer which also includes the sensor barriers to the leads. (c) In the third gate layer, we deposit screening gates and the two central barrier gates $b_{26}$ and $b_{37}$. (d) Finally, we deposit the plunger gates including the sensors. (e) Sketch of the heterostructure and the gate stack on top. The thick black layers in between the gates symbolizes the gates oxides.}
    \label{fig:Design}
\end{figure*}

The device is fabricated on a Ge/Si$_{0.2}$Ge$_{0.8}$ heterostructure with a quantum well buried 55 nm below the surface. The growth is performed by chemical vapor deposition starting from a Si substrate. After growing a thick layer of Ge on the substrate, the Si concentration is linearly increased to reach the desired composition (Fig. \ref{fig:Design}e)\cite{Lodari2021}. The fabrication starts with markers and SiN pads patterned via optical lithography. Subsequently, the ohmic contacts are defined via electron-beam lithography (Fig. \ref{fig:Design}a). The contact material is Pt which, after deposition, is annealed during the oxide deposition (7 nm of ALD grown Al$_2$O$_3$). In the following, we alternate ebeam lithography, metal deposition and oxide growth to define the barrier gate layer (Fig. \ref{fig:Design}b), the screening gate layer (Fig. \ref{fig:Design}c), and the plunger gate layer (Fig. \ref{fig:Design}d). The gate material is Ti/Pd with a thickness of 3/17 nm for the first gate layer, 3/27 nm for the second, and 3/37 for the third gate layer. \\
All measurements are performed in an Oxford Triton dilution refrigerator at a nominal base temperature of 13 mK. We apply magnetic fields in-plane of 10 mT. The device was mounted on a custom-made printed circuit board (PCB). DC voltages from home-built serial peripheral interface (SPI) DAC modules and pulses from a Keysight M3202A arbitrary waveform generator are combined using on-PCB bias tees. RF reflectometry for charge sensing was done using SPI in-phase and quadrature (IQ) demodulation modules and on-PCB LC tank circuits. The demodulated signals were recorded by a Keysight M3102A digitizer.\\

\begin{figure*}[h]
    \centering
    \includegraphics[width=0.5\linewidth]{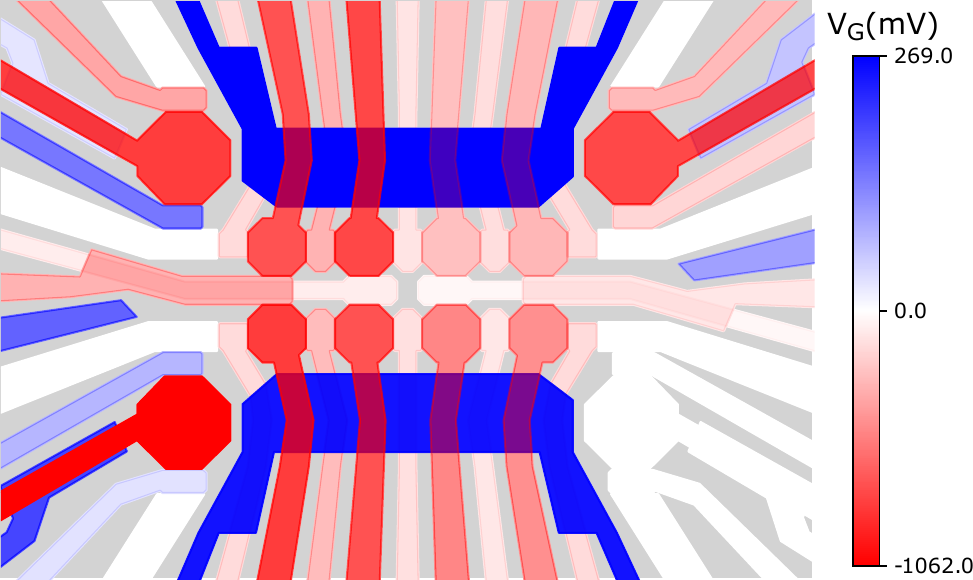}
    \caption{Typical DC voltage configuration in the experiments presented in this work. The color scale reflects the voltage $V_G$ applied to the individual electrodes. Electrodes in white are either grounded or have 0 V applied. All voltages on the bottom right sensor are at 0 V as the sensor plunger was faulty. Starting from the DC configuration, pulses on the AWG channels allow us to quickly change the charge state or the exchange.}
    \label{fig:DC_config}
\end{figure*}
The voltages necessary to tune the device into a regime with a single hole in each of the 8 quantum dots (except dot 2 where 3 holes are confined) and form single-hole transistors (SHTs) in the top left, top right, and bottom left dot are plotted as a heatmap in Fig. \ref{fig:DC_config}. We found that the bottom right sensor was faulty, which is why all voltages in that area are set to 0 V.  

\subsection{Cross-talk matrix for plunger virtualization}
\label{sec:crosstalk matrix}
\begin{figure*}[h]
    \centering
    \includegraphics[width=0.95\textwidth]{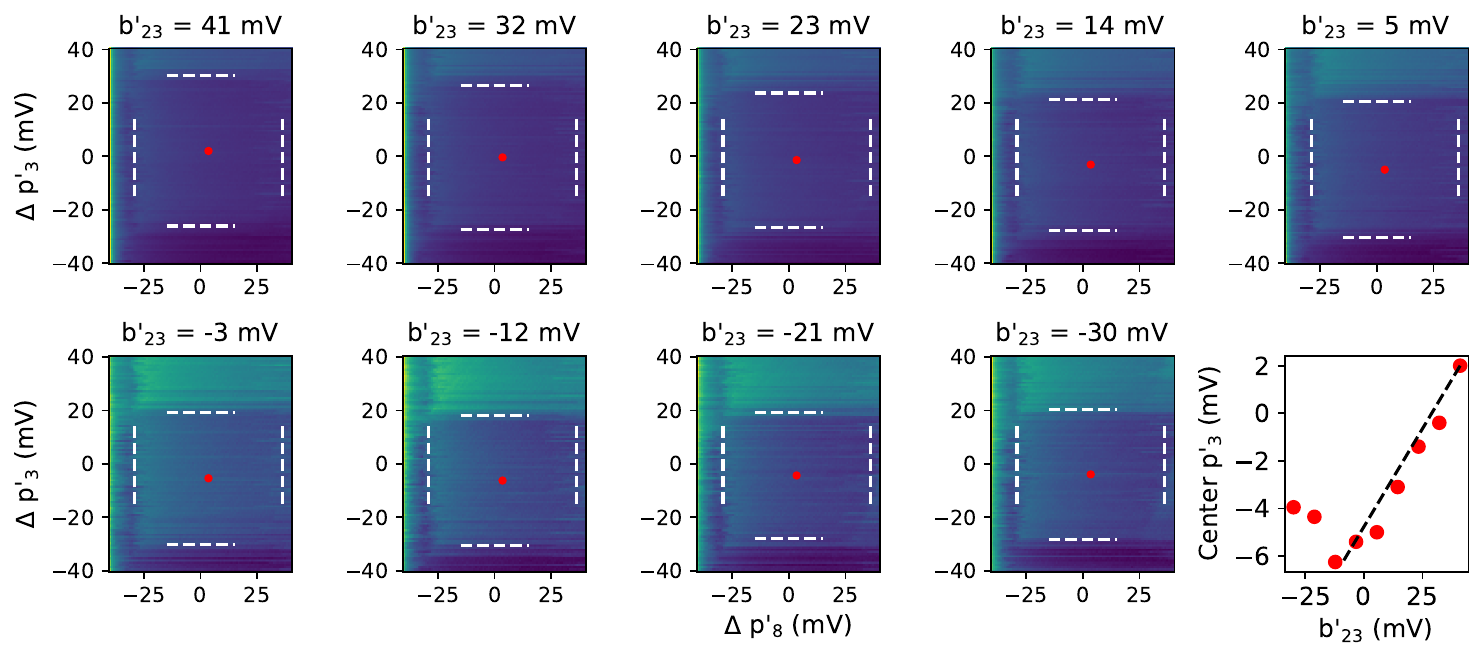}
    \caption{Plunger against barrier virtualization. The plots show charge stability diagrams of $p'_8$ vs $p'_3$ while stepping $b'_{23}$. The white dashed lines mark the charge transitions of dot 8 (vertical) and dot 3 (horizontal). The red dot is the center of the charge stability region which we can follow in all the plots as we lower the value of $b'_{23}$. The last panel in the bottom right shows the extracted center position of $p'_3$ for the different barrier voltages. The black dashed line allows us to extract the crosstalk correction between $b'_{23}$ and $p'_3$. However, we notice that for very negative barrier voltages the center position deviates from this line indicating that the crosstalk has changed. Similar observations were made in reference \cite{Rao2025} and will require more sophisticated non-linear correction schemes.}
    \label{fig:p3vsvb23}
\end{figure*}

\begin{figure*}[h]
    \centering
    \includegraphics[width=0.75\textwidth]{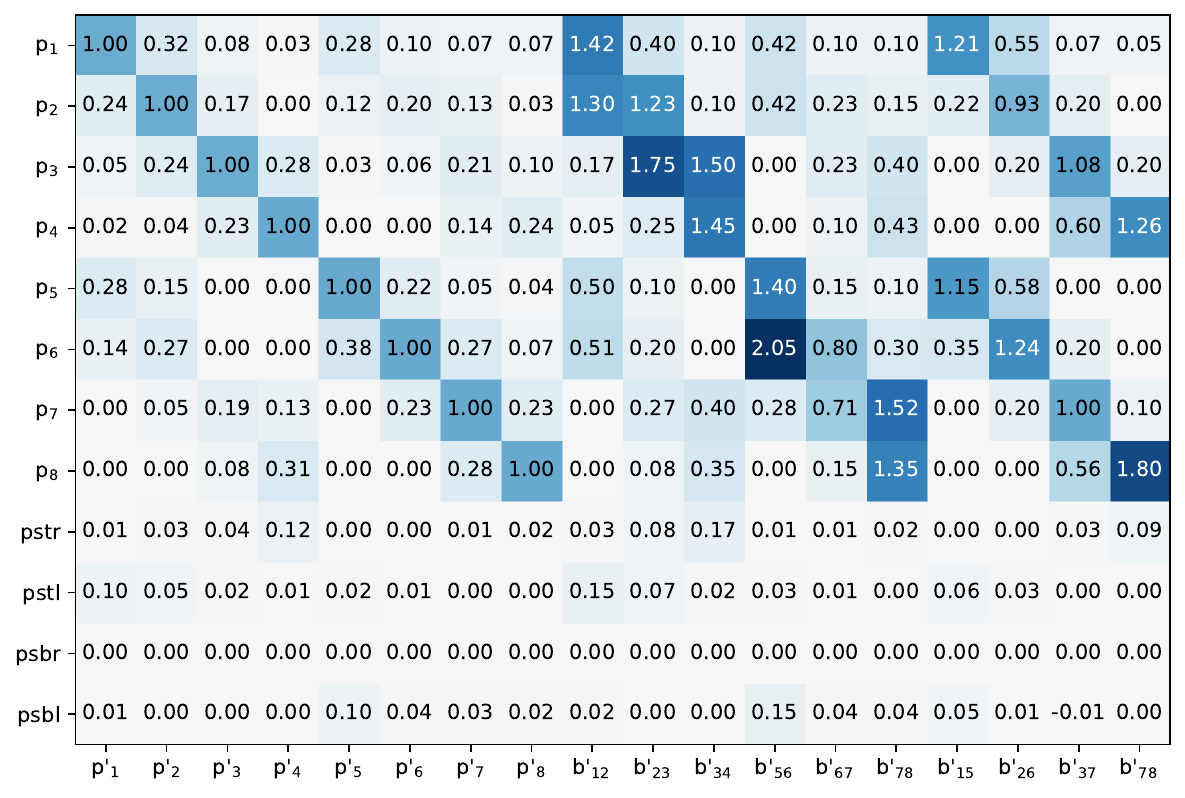}
    \caption{Cross-talk matrix for a first layer of gate virtualization. The linear combination of $p_i$ and
    $b_{ij}$ to orthogonally control the dot potentials is obtained from the inverse of this matrix. "pstr", "pstl", "psbr" and "psbl" refer to the \textbf{p}lunger of \textbf{s}ensor \textbf{t}op ((b)ottom) \textbf{r}ight ((l)eft), respectively. As a consequence of depositing the barrier gates as a first gate layer, large correction pulses on the plungers are needed to compensate for barrier pulses as highlighted by the large cross-talk elements. For example, a pulse on $b'_{56}$ would require a pulse of twice the amplitude on $p_6$ in order to maintain the electrochemical potential of dot 6 unchanged. }
    \label{fig:cross_talk}
\end{figure*}

\begin{figure*}
    \centering
    \includegraphics[width=0.5\linewidth]{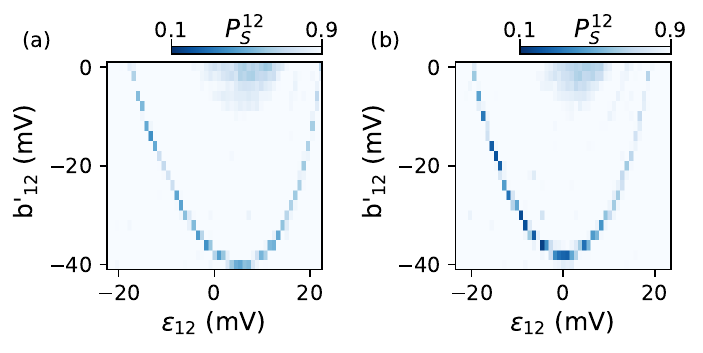}
    \caption{a) ST- avoided crossing of $Q_{12}$ as a function of $\epsilon_{12}$ and $b'_{12}$ with an improper first layer of virtualization leading to a skewed U-shape. 
    b) same measurement as in a) but with corrected virtualization returning a symmetric U-shape.}
    \label{fig:skewed_spin_cup}
\end{figure*}

Virtual plunger gates ease device control and are rather straightforward to obtain from charge stability diagrams. Plunger to plunger cross-talk can be directly extracted from the slopes of the reservoir addition lines. In order to compensated for cross-talk from barriers to plungers it is important to account for the fact that an increased coupling, induced by the barrier, will also modify the plunger to plunger cross-talk element. 
We therefore must ensure we first start from a set of DC gate voltages close to the desired operating conditions. Once a suitable DC voltage configuration is found, we record charge stability diagrams and step the barriers. In this way, it is possible to track the center of the (1,1) charge stability region as a function of the barrier and compensate for this shift (in fact we use a manual version of the method described in \cite{Rao2025}) (Fig. \ref{fig:p3vsvb23}). The resulting virtual gate matrix is depicted in Fig. \ref{fig:cross_talk}. \\
In a next step, we define detunings $\epsilon_{ij} = ap'_j-bp'_i$ and electro-chemical potentials $\mu_{ij} = cp'_j+dp'_{i}$ where $a,b,c,d$ are coefficients that we experimentally determine. If the definitions of $\epsilon_{ij}$ as well as the barrier to plunger virtualization are correct, the $ST^-$ avoided crossing position as a function of detuning and (virtual) barrier should be symmetric U-shaped (in the absence of modulations of the g factor)~\cite{Zhang2024}. An example of an ill-defined virtualization leading to a skewed U-shape is depicted in Fig. \ref{fig:skewed_spin_cup}a, while Fig. \ref{fig:skewed_spin_cup}b shows the same measurement with corrected virtualizations. This step is crucial for the subsequent barrier to barrier compensation, since an unwanted detuning between the quantum dots would enhance the exchange of interest. 

\subsection{Cross-talk extraction}
\begin{figure*}
    \centering
    \includegraphics[width=0.85\linewidth]{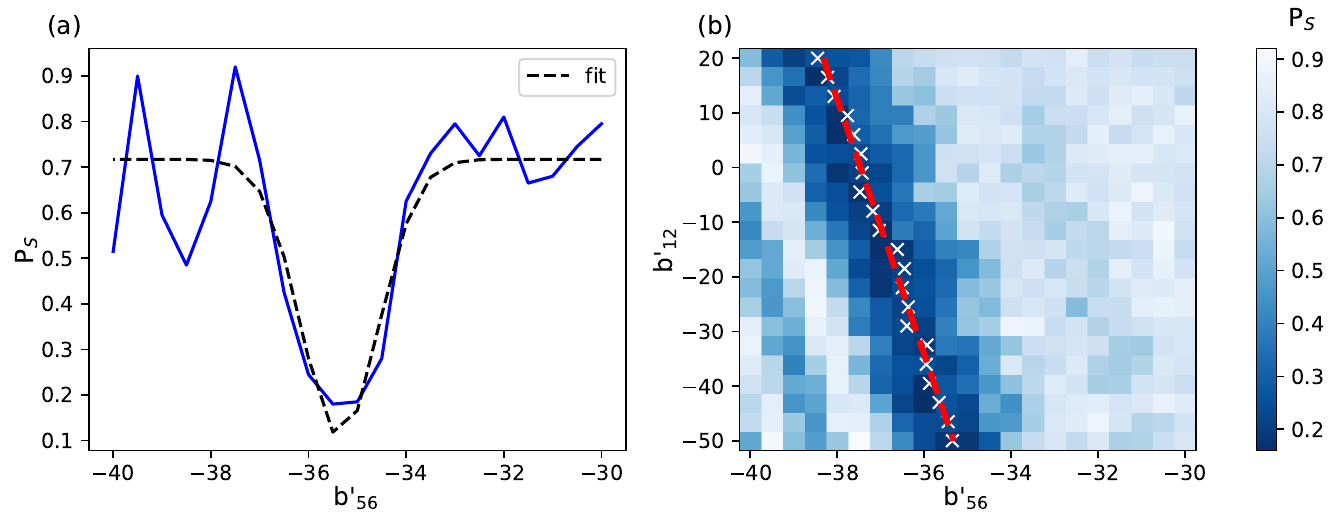}
    \caption{a) Line cut of panel b) corresponding to the top right panel in Fig. \ref{fig:Fig2}}a). we fit a gaussian to the data to extract the position of the minimum in the singlet probability. We repeat this for every value of $b'_{12}$. 
    b) Cross-talk extraction $\alpha_{56}^12$. The white crosses are the extracted minima in $P_S$. The red dashed line is a fit to the data. The slope returns -$\alpha_{56}^12$.
    \label{fig:Cross_talk_extraction}
\end{figure*}

Fig. \ref{fig:Cross_talk_extraction} shows the process for extracting the cross-talk element. In Fig. \ref{fig:Cross_talk_extraction}a we show a linecut of the top leftmost panel of \ref{fig:Fig2}. The black dashed line is a gaussian fit from which we are able to extract the position of the avoided crossing. The white crosses in Fig. \ref{fig:Cross_talk_extraction}b mark the extracted avoided crossing positions to which we can fit the red dashed line. The crosstalk element $\alpha_{56}^12$ is simply the slope multiplied by -1. Such a procedure is repeated for all the other barrier gates to fill out the cross-talk matrix. 

\subsection{Cross-talk for other barriers}
\label{sec:Cross-talk other barriers}
\begin{figure*}
    \centering
    \includegraphics[width=0.95\textwidth]{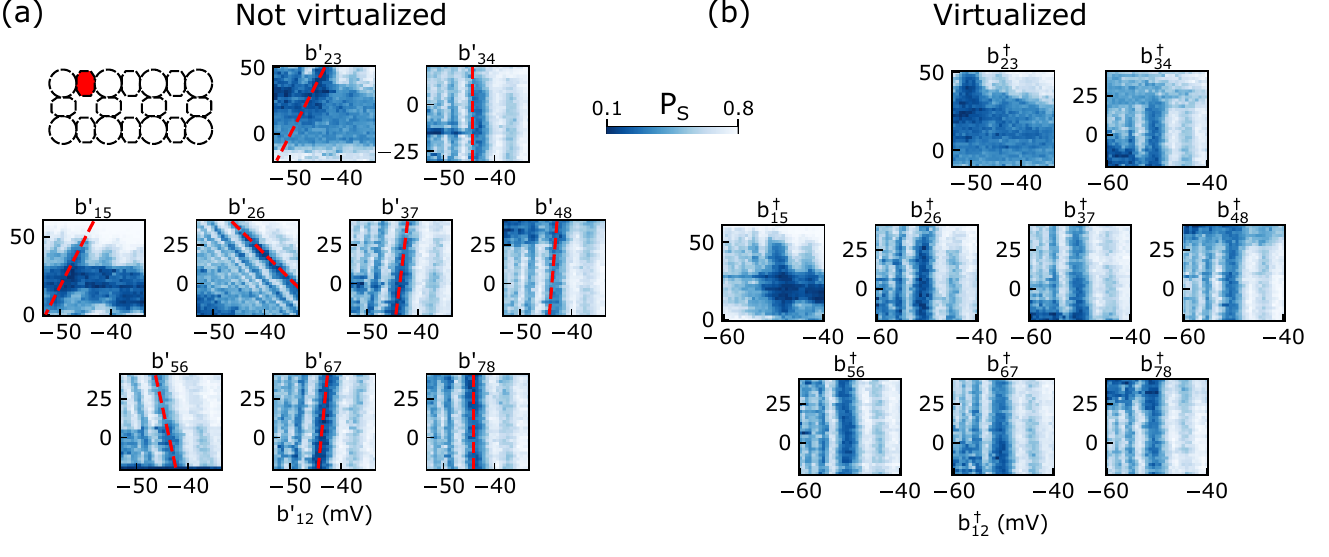}
    \caption{(a) $ST^-$ avoided crossing of $Q_{12}$ as a function of $\mathrm{b'_{12}}$ and all the other barriers $b'_{ij}$. The slopes return the cross-talk element $\alpha_{12}^{mn}$.
    (b) $ST^-$ avoided crossing of $Q_{12}$ as a function of $\mathrm{b^\dagger_{12}}$ and all the other virtual barriers $b^\dagger_{ij}$.}
    \label{fig:12virt}
\end{figure*}

\begin{figure*}
    \centering
    \includegraphics[width=0.95\textwidth]{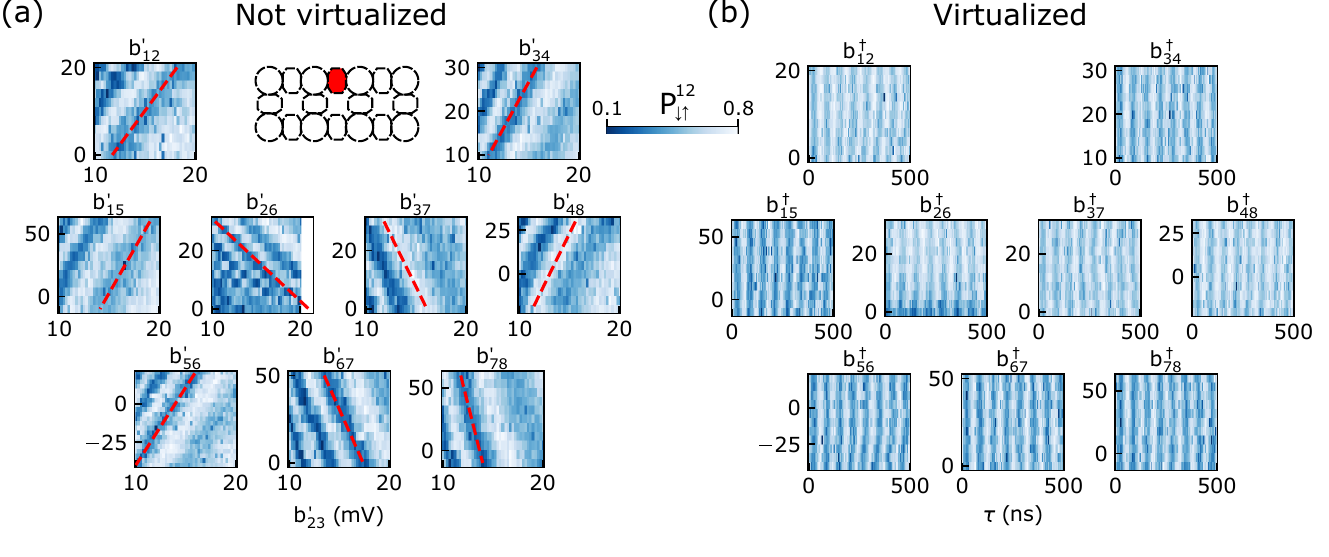}
    \caption{(a) Exchange oscillations of $Q_{12}$ as a function of $b'_{23}$ and all the other barriers $b'_{ij}$. Since we do not have direct access to the $ST^-$ avoided crossing in $Q_{23}$ we opted to initialize $\ket{\downarrow\uparrow\downarrow\downarrow}$ in the top row and record $P_{\downarrow\uparrow}^{12}$. By fixing the evolution time $\tau$ and scanning $b'_{23}$ against the other barriers we are still able to follow features and extract the cross-talk element, although no feature stands out more than others.
    (b) Exchange oscillations of $Q_{12}$ as a function of $b^\dagger_{mn}$ and duration $\tau$. None of the virtual barriers $b^\dagger_{ij}$ affect the oscillation frequency which indicates that the cross-talk to $b^\dagger_{23}$ is correctly compensated.}
    \label{fig:23virt}
\end{figure*}

\begin{figure*}
    \centering
    \includegraphics[width=0.95\textwidth]{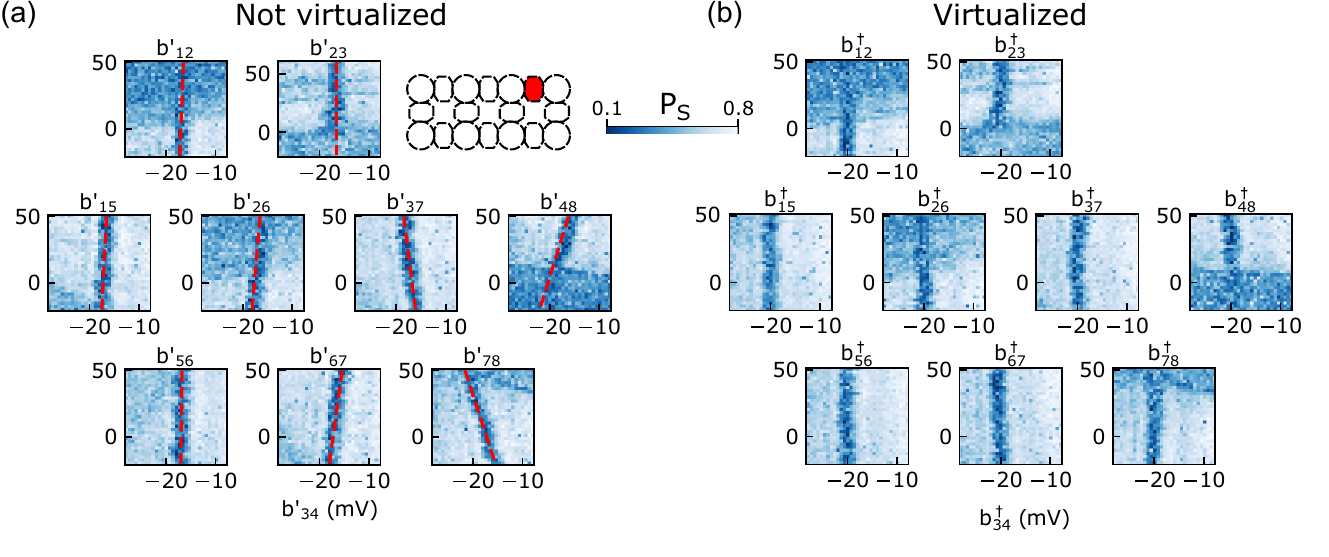}
    \caption{(a) $ST^-$ avoided crossing of $Q_{34}$ as a function of $\mathrm{b'_{34}}$ and all the other barriers $b'_{ij}$. The slopes return the cross-talk element $\alpha_{34}^{mn}$.
    (b) $ST^-$ avoided crossing of $Q_{34}$ as a function of $\mathrm{b^\dagger_{34}}$ and all the other virtual barriers $b^\dagger_{ij}$.
    }
    \label{fig:34virt}
\end{figure*}

\begin{figure*}
    \centering
    \includegraphics[width=0.95\textwidth]{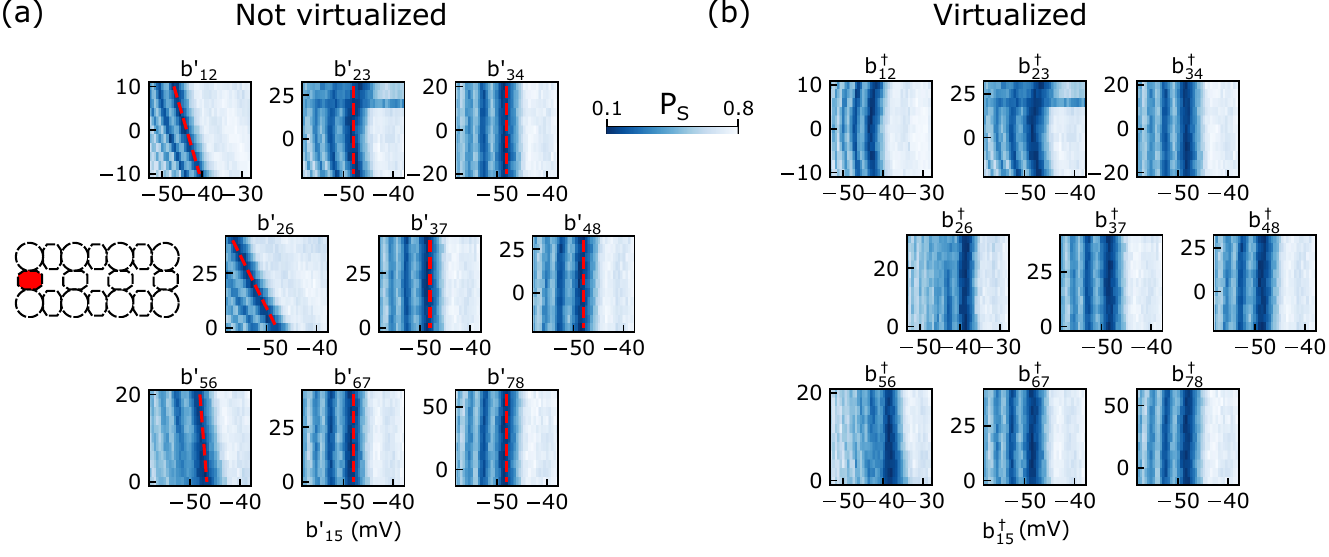}
    \caption{(a) $ST^-$ avoided crossing of $Q_{15}$ as a function of $\mathrm{b'_{15}}$ and all the other barriers $b'_{ij}$. The slopes return the cross-talk element $\alpha_{15}^{mn}$.
    (b) $ST^-$ avoided crossing of $Q_{15}$ as a function of $\mathrm{b^\dagger_{15}}$ and all the other virtual barriers $b^\dagger_{ij}$. Except for $b^\dagger_{23}$ all the virtualizations are correct. We note that when a gate cross-talk element is zero $b'_{mn}=b^\dagger_{mn}$ for this case and we simply reproduced the plots from (a) also in (b).}
    \label{fig:15virt}
\end{figure*}

\begin{figure*}
    \centering
    \includegraphics[width=0.95\textwidth]{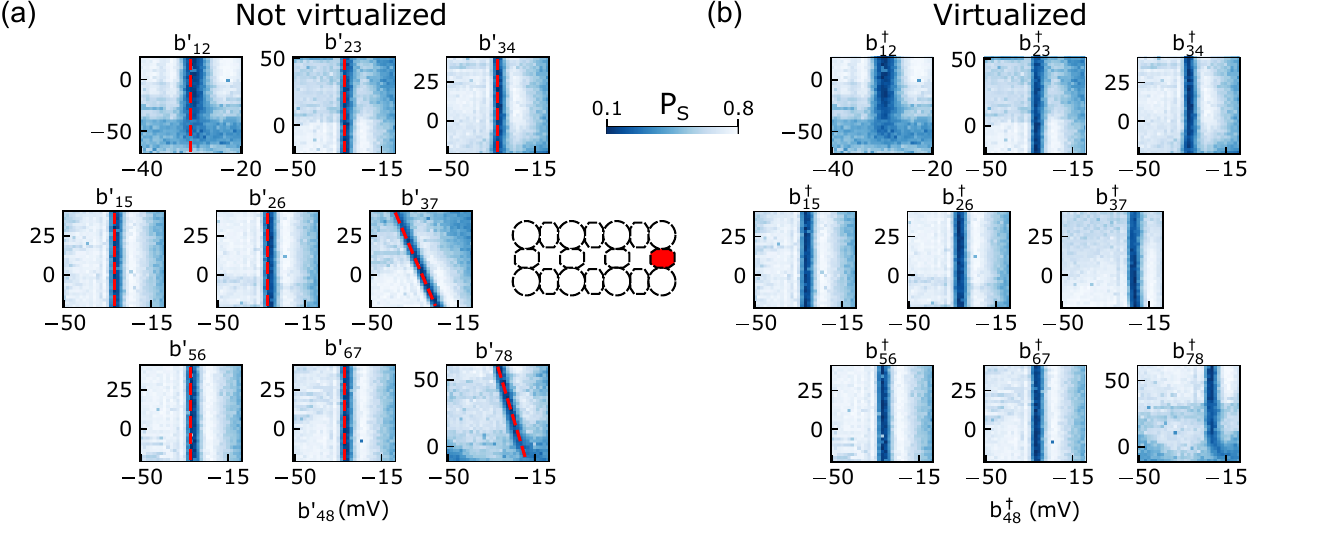}
    \caption{(a) $ST^-$ avoided crossing of $Q_{48}$ as a function of $\mathrm{b'_{48}}$ and all the other barriers $b'_{ij}$. The slopes return the cross-talk element $\alpha_{48}^{mn}$.
    (b) $ST^-$ avoided crossing of $Q_{48}$ as a function of $\mathrm{b^\dagger_{48}}$ and all the other virtual barriers $b^\dagger_{ij}$. We note that when a gate cross-talk element is zero $b'_{mn}=b^\dagger_{mn}$ for this case and we simply reproduced the plots from (a) also in (b).}
    \label{fig:48virt}
\end{figure*}

\begin{figure*}
    \centering
    \includegraphics[width=0.95\textwidth]{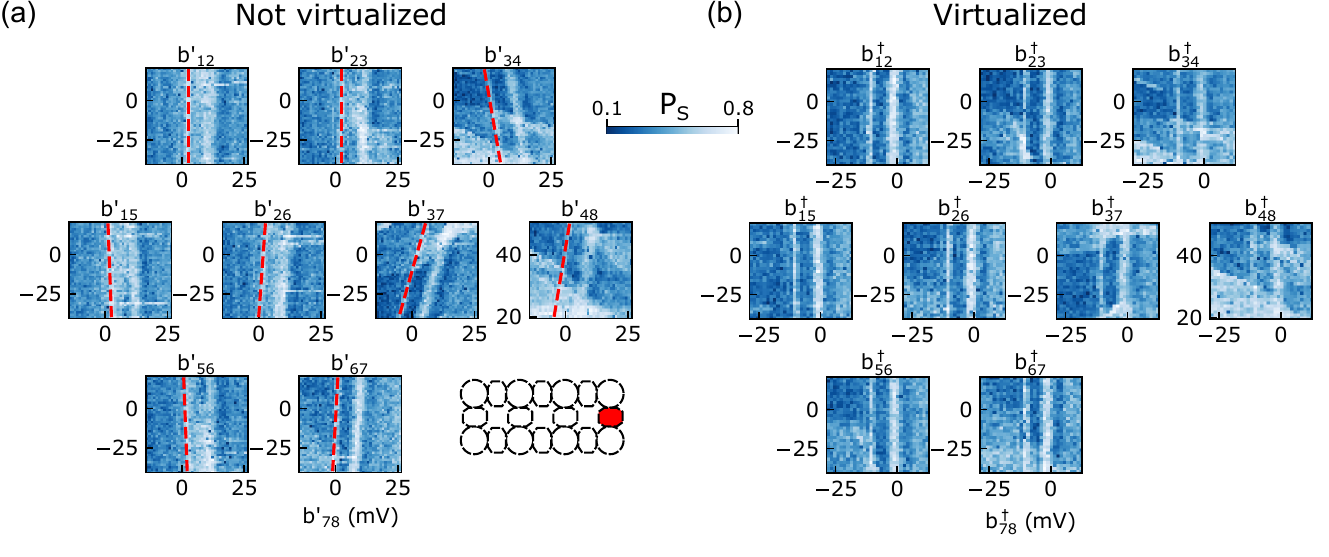}
    \caption{(a) $ST^-$ avoided crossing of $Q_{78}$ as a function of $\mathrm{b'_{78}}$ and all the other barriers $b'_{ij}$. The slopes return the cross-talk element $\alpha_{78}^{mn}$.
    (b) $ST^-$ avoided crossing of $Q_{78}$ as a function of $\mathrm{b^\dagger_{78}}$ and all the other virtual barriers $b^\dagger_{ij}$.}
    \label{fig:78virt}
\end{figure*}

Fig. \ref{fig:12virt}, \ref{fig:23virt}, \ref{fig:34virt}, \ref{fig:15virt}, \ref{fig:48virt} and Fig. \ref{fig:78virt} present data analogous to the data shown in Fig. \ref{fig:cross_talk} for $Q_{56}$. Specifically, they show the avoided crossing features or exchange oscillations of $Q_{12}$, $Q_{23}$, $Q_{34}$, $Q_{15}$, $Q_{48}$ and $Q_{78}$, respectively. In all the figures, the plots associated with the respective barriers are organized to reflect the geometry of the device. We always chose the ranges in a way to not induce any exchange for nearest neighbor barriers. For $Q_{23}$ we do not have access to direct PSB readout. Therefore, we opted to initialize the state $\ket{\downarrow\uparrow\downarrow\downarrow}$ in the top row and record exchange oscillations at a fixed duration $\tau$ and scan $b'_{23}$ against $b'_{mn}$. We record the probability $P_{\downarrow\uparrow}^{12}$ which, with $b'_{12}$ and $b'_{34}$ not inducing any exchange, oscillates at a frequency determined by $b'_{23}$. While the resulting features are not as clear and isolated as the ones from the avoided crossing, they still allow us to extract a cross-talk element. Fig. \ref{fig:23virt}b shows exchange oscillations between spins 2 and 3, again recorded as $P_{\downarrow\uparrow}^{12}$ which are not influenced by any of the $b^\dagger_{mn}$ demonstrating that cross-talk is compensated.\\
Lastly, we note that the cross-talk elements reported in Fig. \ref{fig:AllVirtAndMatrix}b are the ones we used in the experiments in Fig. \ref{fig:ResonantConditions} and are not necessarily the same as the slopes in the measurements here would suggest. 

\subsection{Initialization and read-out schemes}
\begin{figure*}[h]
    \centering
    \includegraphics[width=\textwidth]{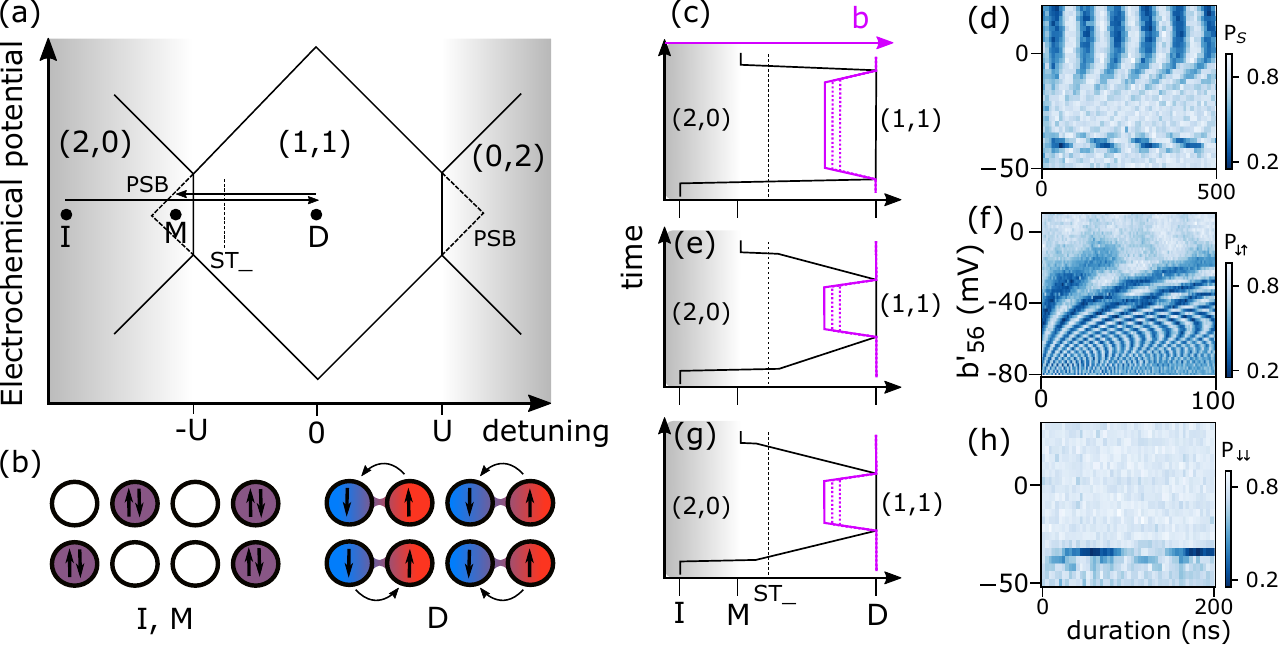}
    \caption{(a) Schematic of a charge stability diagram of a double quantum dot as a function of detuning and the electro-chemical potentials of the two dots, which are uniformly varied. When the detuning is equal to the charging energy $U$, charges get transferred between the dots through the vertical transition lines. The region inside the dashed triangle corresponds to the metastable region where Pauli-spin-blockade occurs assuming the triplet excited state falls outside the triangle (otherwise the triangle gets truncated). A typical experiment starts with two charges in one of the two dots ((2,0) or (0,2) charge region, as depicted in the left sketch in (b)). We then pulse the system into the (1,1) region at the dwell point D. Depending on the ramp type ((c), (e) or (g)) we initialize $\ket{S}$, $\ket{\downarrow\uparrow}$ or $\ket{\downarrow\downarrow}$. After letting the system evolve we pulse back to the measurement point M and perform single-shot readout of the final state.
    (b) Sketch of the typical charge and spin configuration at the initialization (I) and measurement (M) point (left) as well as at the dwell point (D) (right). 
    (c) Pulse scheme to obtain $ST^0$ oscillations when initializing in S and reading S (by ``reading a state'', we mean distinguishing this state from the other three states in the two-qubit space). After pulsing the detuning quickly to D we diabatically pulse on the barrier voltage and next diabatically pulse the detuning back to M. A typical oscillation pattern as in (d) emerges.
    (d) Singlet-triplet oscillations of $Q_{56}$ as a function of $b'_{56}$. For more positive barrier voltages we clearly see $S-T^0$ oscillations at a frequency $hf=\Delta g_{56} \mu_B B$. As the barrier voltage is lowered and exchange increases, the oscillation frequency increases and the visibility is lowered indicating the initial $\ket{S}$ is now an eigenstate of the system. Around $b'_{56} = \SI{-40}{\milli\volt}$ another oscillation can be observed corresponding to $S-T^-$ oscillations at the $ST^-$ avoided crossing. 
    (e) Pulse scheme to obtain exchange oscillations for initialization and readout of $\ket{\uparrow\downarrow}$. By ramping adiabatically with respect to $\Delta g \mu_B B$ (after diabatically sweeping over the $ST^-$ avoided crossing) we are able to initialize an antiparallel spin state at the dwell point. A diabatic pulse on the barrier will then induce SWAP oscillations between $\ket{\uparrow\downarrow}$ and $\ket{\downarrow\uparrow}$. Ramping the detuning back adiabatically until the avoided crossing and then diabatically until the measurement point takes the final $\ket{\uparrow\downarrow}$ state onto a singlet (2,0) during readout effectively returning $P_{\uparrow\downarrow}$.  
    (f) Exchange oscillations as a function of $b'_{56}$ utilizing the pulse scheme in (e). This time we see the amplitude of the oscillations increase as we lower the barrier voltage and exchange is enhanced, as expected for SWAP oscillations starting from $\ket{\downarrow\uparrow}$. 
    (g) Pulse scheme to obtain oscillations when initializing and reading $\ket{T^-}$ by adiabatically ramping over the $ST^-$ avoided crossing. The same adiabatic ramp before read-out takes $\ket{T^-}$ onto a singlet (2,0) during read-out effectively returning $P_{\downarrow\downarrow}$. Between initialization and readout, a diabatic barrier pulse is applied.   
    (h) With the pulse scheme in (g) we now only see oscillations at the $ST^-$ avoided crossing as expected when starting from $\ket{T^-}$. }
    \label{fig:Init_read}
\end{figure*}

Fig. \ref{fig:Init_read} schematically shows the different initialization and read-out schemes. A fast pulse (Fig. \ref{fig:Init_read}c) initializes and reads $\ket{S}$, a ramped pulse starting after the avoided crossing (Fig. \ref{fig:Init_read}e) initializes $\ket{\downarrow\uparrow}$ or $\ket{\uparrow\downarrow}$, while a ramped pulse starting before the avoided crossing initializes $\ket{\downarrow\downarrow}$. During the time evolution in (1,1) we pulse on a barrier to induce exchange. Combining the initialization with an appropriate pulse shape to the read-out point unveils $ST^0$ oscillations (Fig. \ref{fig:Init_read}d), exchange oscillations (Fig. \ref{fig:Init_read}f) or $ST^-$ oscillations (Fig. \ref{fig:Init_read}h). 
\label{sec:Init and read}

\subsection{g-factors}
\label{sec:g-factors}
\begin{figure*}[h]
    \centering
    \includegraphics[width=0.5\textwidth]{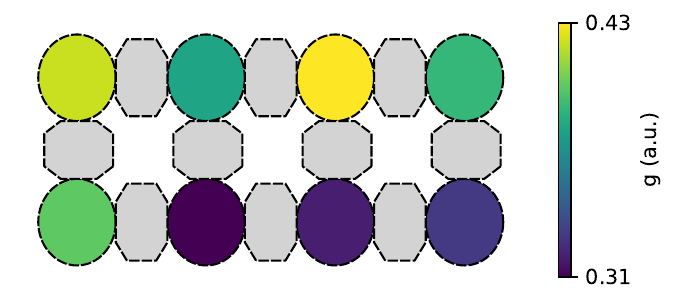}
    \caption{G-factors for the spins in the different dots. The g-factors were extracted from electric dipole spin resonance experiments (data not shown).}
    \label{fig:g-factors}
\end{figure*}

We extract the resonance frequencies $f_i$ of our spins by means of electric dipole spin resonance (EDSR) and find our effective g-factors as $g_i=\frac{f_i}{\mu_B B}h$ at a field of $B = \SI{10}{\milli\tesla}$. $h$ is Plank's constant.  The results are summarized in Fig. \ref{fig:g-factors}. 

\subsection{Exchange profiles for barrier gates}
\label{sec:J-profiles}
\begin{figure*}[h]
    \centering
    \includegraphics[width=0.9\textwidth]{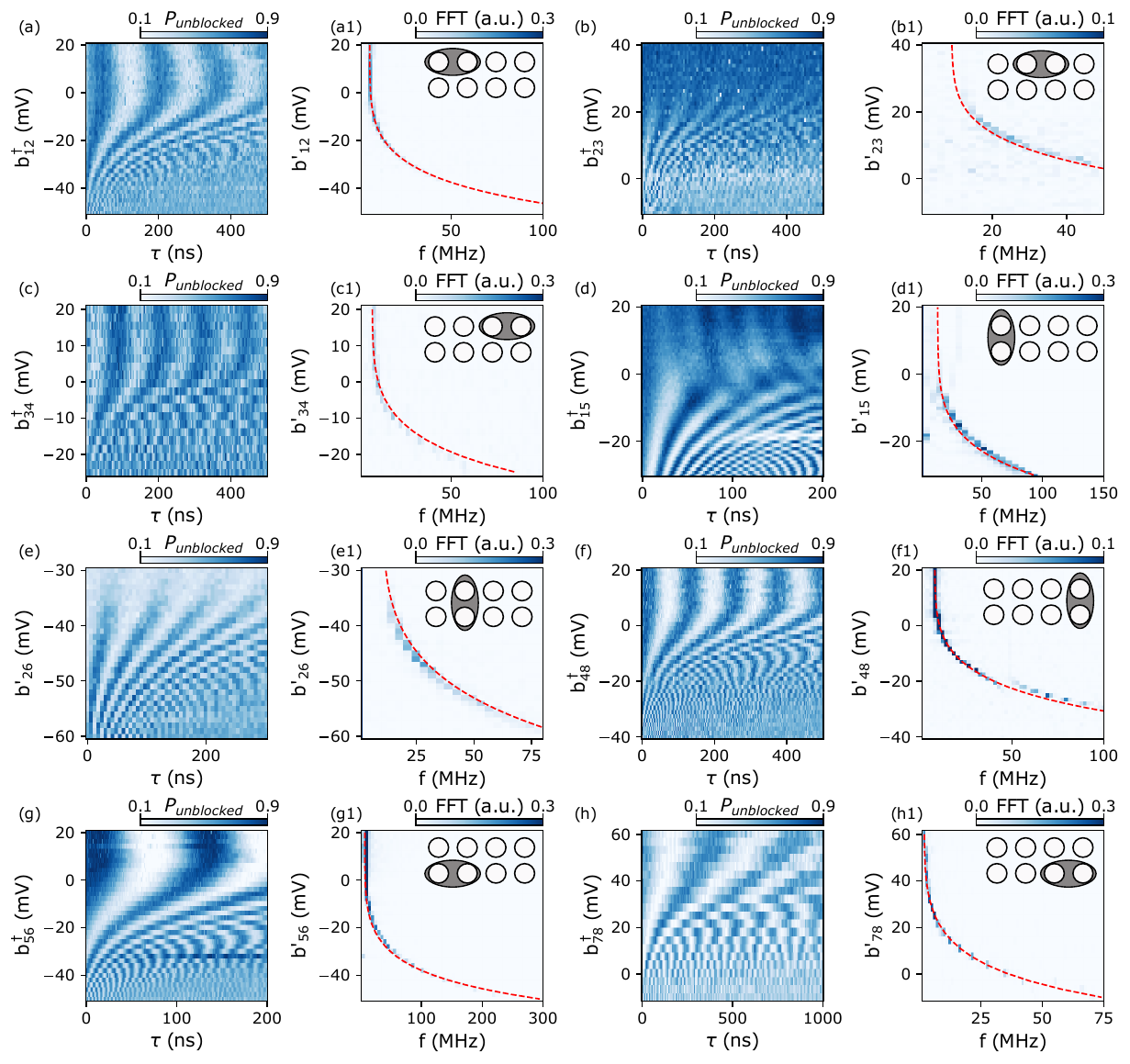}
    \caption{(a)-(h) Exchange oscillations as a function of the barriers. (a1)-(h1) FFTs of the oscillations and fit to the exchange formula. The insets schematically show which ST qubit is activated. All other exchanges are switched off. }
    \label{fig:exch_profiles}
\end{figure*}

We typically operate every singlet-triplet qubit at its symmetry point ($\epsilon_{ij} = 0$). Here the exchange is determined solely by the height of the tunnel barrier ($J = \frac{2t^2}{U}$). It is common practice to approximate the exchange dependence on the barrier by an exponential function $J(b_{ij}) = J_{0}\exp\left(k(b_{ij}-b_{0ij})\right)$ where $b_{0ij}$ is an offset determined by the DC voltage configuration and $J_0=\SI{1}{\mega\hertz}$. $k$ can be interpreted as a barrier lever arm, e.g. how strongly the barrier voltage affects the exchange. This is dependent on the gate layout and the DC voltage configuration. Fig. \ref{fig:exch_profiles} shows the exchange profiles of the singlet triplet qubits we measured alongside their FFTs and fits the to the exchange formula. Table \ref{tab:Exch_summary} summarizes the extracted values for all the singlet triplet qubits measured.

\begin{table}[]
    \centering
    \begin{tabular}{|c|c|c|}
        \hline
        \textbf{gate}  & $k (\SI{}{\milli\volt^{-1}})$ & $\mathrm{b^\dagger_0} (\SI{}{\milli\volt})$\\
        \hline
        $\mathrm{b^\dagger_{12}}$ & -79.29 & 11.7 \\ 
        \hline
        $\mathrm{b^\dagger_{23}}$ &  -94.72 & 44.1 \\ 
        \hline
        $\mathrm{b^\dagger_{34}}$ &  -96.69 & 21.0\\ 
        \hline
        $\mathrm{b^\dagger_{15}}$ &  -90.05 & 20.2\\ 
        \hline
        $\mathrm{b'_{26}}$ &  -90.02 & -9.8\\ 
        \hline
        $\mathrm{b'_{37}}$ &  n.m & n.m \\ 
        \hline
        $\mathrm{b^\dagger_{48}}$ &  -86.38 & 22.6 \\ 
        \hline
        $\mathrm{b^\dagger_{56}}$ &  -90.00 & 13.2\\ 
        \hline
        $\mathrm{b'_{67}}$ &  n.m& n.m\\ 
        \hline
        $\mathrm{b^\dagger_{78}}$ &  -71.6 & 50.1\\ 
        \hline
    \end{tabular}
    \caption{Summary of the measured exchange dependences on the barrier gates. $b'_{37}$ and $b'_{67}$ have not been measured and that $b'_{26}$ is not virtualized. The DC voltages ensure that $|b^\dagger_{0,ij}|<\SI{50}{\milli\volt}$, enabling a large on-off ratio within a comfortable pulse amplitude for the AWG. }
    \label{tab:Exch_summary}
\end{table}

\subsection{Singlet-triplet resonant conditions}
In this section we want to further elucidate the reasoning behind the resonant 4-spin $ST^-$ and $ST^0$ conditions which are plotted in Fig. \ref{fig:ResonantConditions} of the main text.  
In the presence of an external magnetic field $B$ and isotropic but site dependent g-factors $g_i$, with $\hbar = 1$, the Heisenberg Hamiltonian assumes the form:
\begin{equation}
    H = \sum_{i}g_i\mu_B B S_{z,i}+\sum_{i}\Delta_{SO,i} S_{x,i}+\sum_{<i,j>} J_{ij}\left( \bm{S_iS_j}-\frac{1}{4}\right)
\label{eq:4spinHam_supp}
\end{equation}
where $\mathbf{S} = (S_x, S_y, S_z)$ is the spin operator on site $i$.  Here, $\Delta_{SO}$ consists of anisotropic g-tensor components and the intrinsic spin-orbit interaction in the spin-orbit frame~\cite{Geyer2024}. The influence of additional $S_y$ terms in the spin-orbit term can neglected if we only analyze the isolated $ST^-$ ($ST^0$) subspace~\cite{Zhang2024}.
For four spins $i,j,k,l$ it is instructive to write the basis states of the different spin subspaces in the familiar singlet-triplet basis of two-spin states. These are summarized in table \ref{tab:Heis_states}. Without the spin-orbit interaction, a spin system initialized in one of these subspaces should evolve only within that subspace, e.g. the total spin number is conserved. We particularly want to draw the attention to the two lowest states in the $T^0$ and $T^-$ subspace. These states contain one singlet and one triplet each.

\label{sec:ST resonant cond}
\begin{table}[]
    \centering
    \begin{tabular}{|c|c|c|}
    \hline
       Subspace  & (S, $m_S$) & two-qubit basis states \\
    \hline
         & $(2,\pm 2)$ & $\ket{T_{ij}^{\pm}T_{kl}^{\pm}}$ \\
       $Q$  & $(2,\pm 1)$ & $\frac{1}{\sqrt{2}}(\ket{T_{ij}^{0}T_{kl}^{\pm}}+\ket{T_{ij}^{\pm}T_{kl}^{0}})$ \\
         & $(2,0)$ & $\frac{1}{\sqrt{6}}(\ket{T_{ij}^{+}T_{kl}^{-}}+\ket{T_{ij}^{-}T_{kl}^{+}}+2\ket{T_{ij}^{0}T_{kl}^{0}})$ \\
    \hline 
        & & $\frac{1}{\sqrt{2}}(\ket{T_{ij}^0 T_{kl}^{\pm}}-\ket{T_{ij}^{\pm} T_{kl}^{0}})$\\
    $T^{\pm}$ & $(1, \pm 1)$ & $\ket{T_{ij^{\pm}}S_{kl}}$ \\
        & & $\ket{S_{ij}T_{kl}^{\pm}}$\\
    \hline 
        & & $\frac{1}{\sqrt{2}}(\ket{T_{ij}^+ T_{kl}^{-}}-\ket{T_{ij}^{-} T_{kl}^{+}})$\\
    $T^{0}$ & $(1, 0)$ & $\ket{T_{ij}^{0} S_{kl}}$ \\
        & & $\ket{S_{ij}T_{kl}^{0}}$\\
    \hline
     $S$ & $(0,0)$ & $\frac{1}{\sqrt{3}}(\ket{T_{ij}^+T_{kl}^-}+\ket{T_{ij}^-T_{kl}^+}-\ket{T_{ij}^0 T_{kl}^0})$ \\
         & & $\ket{S_{ij}S_{kl}}$\\
    \hline       
    \end{tabular}
    \caption{Four-spin shared eigenstates of $\hat{S}^2$ and $\hat{S}^z$ expressed in a basis of two-spin singlets and triplets. As the Heisenberg Hamiltonian is spin conserving, it only couples states within the same subspace. Note that these states are, in general, not eigenstates of the Heisenberg Hamiltonian. The lowest two states  of the $T^0$ and $T^-$ subspace in the table are used to find resonant four spin conditions in Fig. \ref{fig:ResonantConditions}.}
    \label{tab:Heis_states}
\end{table}

\subsubsection{$\ket{ST^-}, \ket{T^-S}$ subspace}
If we reduce ourselves to the basis $\bigl\{ \ket{S_{ij}S_{kl}}, \ket{S_{ij}T^-_{kl}}, \ket{T^-_{ij}S_{kl}}, \ket{T^-_{ij}T^-_{kl}}\bigr\}$ the Hamiltonian takes the form \cite{Zhang2024}: 
\begin{equation*}
    H_{ST^-}= \left( 
    \begin{matrix}
    -J_{ij}-J_{kl} & \frac{\Delta_{SO, kl}}{2} & \frac{\Delta_{SO, ij}}{2} & 0 \\
    \frac{\Delta_{SO, kl}}{2} & -J_{ij}-\overline{E}_{z,kl} & -\frac{J_{jk}}{4} & \frac{\Delta_{SO, ij}}{2} \\
    \frac{\Delta_{SO, ij}}{2} & -\frac{J_{jk}}{4} & -\overline{E}_{z,ij}-J_{kl} & \frac{\Delta_{SO, kl}}{2} \\
    0 & \frac{\Delta_{SO, ij}}{2} & \frac{\Delta_{SO, kl}}{2} & -2\overline{E}_{z,ijkl}+\frac{J_{jk}}{4}
    \end{matrix}\right)
    \label{eq:HST-}
\end{equation*}
Compared to eq. \ref{HQ} of the main text we have included the spin-orbit part as well which gives rise to leakage terms outside of the $\ket{S_{ij}T^-_{kl}}, \ket{T^-_{ij}S_{kl}}$ subspace. However, these occur only at the respective avoided crossings ($J_{ij}=\overline{E_Z}_{ij}$) and it is easy to operate away from these locations. When $|J_{ij}-\overline{E}_{Z,ij}|=|J_{kl}-\overline{E}_{Z,kl}|$ the middle two diagonal terms are equal and the off diagonal terms $\frac{J_{jk}}{4}$ become dominant. Initializing $\ket{S_{ij}T^-_{kl}}$ and pulsing quickly to this condition will induce $\ket{S_{ij}T^-_{kl}}\leftrightarrow \ket{T^-_{ij}S_{kl}}$ oscillations with a frequency $f_{ST^-} = \frac{J_{jk}}{h}$ \cite{Wang2023}.

\subsubsection{$\ket{ST^0}, \ket{T^0S}$ subspace}
\begin{figure*}
    \centering
    \includegraphics[width=0.85\textwidth]{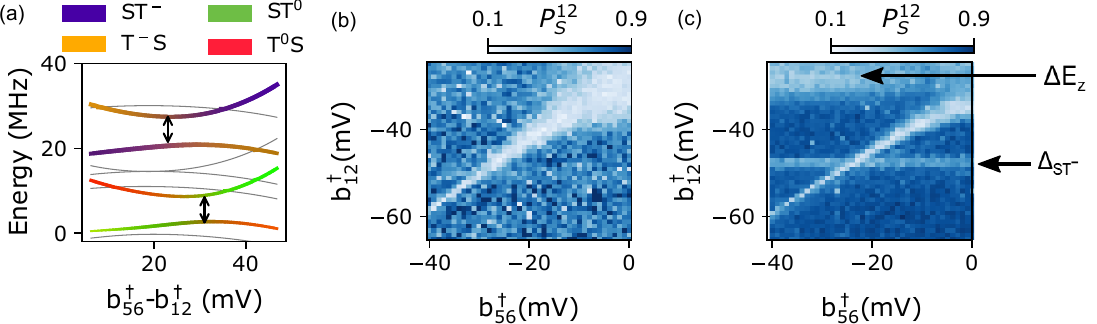}
    \caption{(a) Simulated energy diagram of the 4-spin system 2-1-5-6  as a function of $b^\dagger_{56}-b^\dagger_{12}$ as scanned in Fig. \ref{fig:ResonantConditions}b. The overlap with the 4 important states in the legend is color and thickness coded. We can observe 2 avoided crossings, one between $\ket{ST^-}\leftrightarrow\ket{T^-S}$ and another for $\ket{ST^0}\leftrightarrow\ket{T^0S}$. The fact that the avoided crossings occur at different gate voltages reflects the slightly different requirements for $J_{12}$ and $J_{56}$. However, the size of both avoided crossings is equal and solely determined by $J_{15}$. 
    (b) same as Fig. \ref{fig:ResonantConditions}a.
    (c) Resonant $\ket{S_{12}T^0_{56}}$ condition as a function of $b^\dagger_{56}$ and $b^\dagger_{12}$. We initialize $\ket{S_{12}T^0_{56}}$ and let the system evolve for 100 ns at each voltage point with a small exchange opened through $b^\dagger_{15}$. Like in (b) we can identify a sharp resonant condition which occurs at slightly different gate voltages than (b) as $\ket{S_{12}T^0_{56}}$ evolves to $\ket{T^0_{12}S_{56}}$. In the bottom left quadrant, however, the two resonant conditions approach each other. This is expected since $J_{12}$ and $J_{56}$ become the dominant energies in the system. We can also identify two leakage features indicated by the black arrows. One pertains to the singlet triplet avoided crossing of $Q_{12}$ ($\Delta_{SO}$), while the other, at more positive voltages of $b^\dagger_{12}$ can be attributed to $ST^0$ oscillations in $Q_{12}$ since $J_{12}\approx \Delta E_{Z,12}$. }
    \label{fig:1256_Energy_diagram}
\end{figure*}

If we reduce ourselves to the basis $\bigl\{ \ket{S_{ij}S_{kl}}, \ket{S_{ij}T^0_{kl}}, \ket{T^0_{ij}S_{kl}}, \ket{T^0_{ij}T^0_{kl}}\bigr\}$ the Hamiltonian takes the form: 

\begin{equation*}
    H_{ST^0}= \left( 
    \begin{matrix}
    -J_{ij}-J_{kl} & \Delta E_{Z,kl} & \Delta E_{Z,ij} & -\frac{J_{jk}}{4}\\
    \Delta E_{Z,kl}& -J_{ij} & -\frac{J_{jk}}{4} & \Delta E_{Z,ij} \\
    \Delta E_{Z,ij} & -\frac{J_{jk}}{4}& -J_{kl} & \Delta E_{Z,kl} \\
    -\frac{J_{jk}}{4} & \Delta E_{Z,ij}& \Delta E_{Z,kl} & 0
    \end{matrix}\right)
    \label{eq:HST0}
\end{equation*}
Similar to the Hamiltonian \ref{eq:HST-}, we can identify a resonant condition: $\sqrt{J_{ij}^2+\Delta E_{Z,ij}^2}= \sqrt{J_{kl}^2+\Delta E_{Z,kl}^2}$. Again we will find that at these special conditions we can induce $\ket{S_{ij}T^0_{kl}}\leftrightarrow \ket{T^0_{ij}S_{kl}}$ oscillations at a frequency $f_{ST^0} = \frac{J_{jk}}{h}$. In this case, however, the leakage terms outside this subspace are given by $\Delta E_{Z,ij,kl}$ and are slightly more difficult to avoid. In fact, we need to ensure that $J_{ij}\gg \Delta E_{Z,ij}$ and $J_{kl}\gg \Delta E_{Z,kl}$ which, for the top right quadrants in Fig. 4a,e,i is not always given. These notions are summarized in Fig. \ref{fig:1256_Energy_diagram}. In Fig. \ref{fig:1256_Energy_diagram}a we plot the energy diagram of the 4-spin system 2-1-5-6  as a function of $b^\dagger_{56}-b^\dagger_{12}$ as scanned in Fig. \ref{fig:ResonantConditions}b with a small exchange between spins 1 and 5 in the chain induced by $b^\dagger_{15}$. The overlap with the 4 important states in the legend is color and thickness coded. We can observe 2 avoided crossings, one between $\ket{ST^-}\leftrightarrow\ket{T^-S}$ and another for $\ket{ST^0}\leftrightarrow\ket{T^0S}$. The fact that the avoided crossings occur at different gate voltages reflects the slightly different requirements for $J_{12}$ and $J_{56}$. However, the size of both avoided crossings is equal and solely determined by $J_{15}$. \\
Finally, we point out that since the read-out in PSB is not capable of distinguishing $\ket{T^0}$ from $\ket{T^-}$ we will find more 'leakage' features that, while not directly coupling the different subspaces, will lead to deviations from the expected resonant condition positions. This can be appreciated in Fig. \ref{fig:1256_Energy_diagram}c where we record the $S_{12}T_{56}^0$ resonant condition. Apart from the expected diagonal feature, we also observe two horizontal features indicated by the arrows. One can be attributed to $ST^0$ oscillations in $Q_{12}$, the other to $ST^-$ oscillations at the spin-orbit anticrossing of $Q_{12}$. Apart from the spin-orbit induced leakage terms, the Hamiltonian which we use to simulate the system contains all the necessary information to reproduce the experimental results. 

\subsection{Additional data for 1256 resonant condition}
\label{sec:additional_1256}

\begin{figure*}[h]
    \centering
    \includegraphics[width=0.85\textwidth]{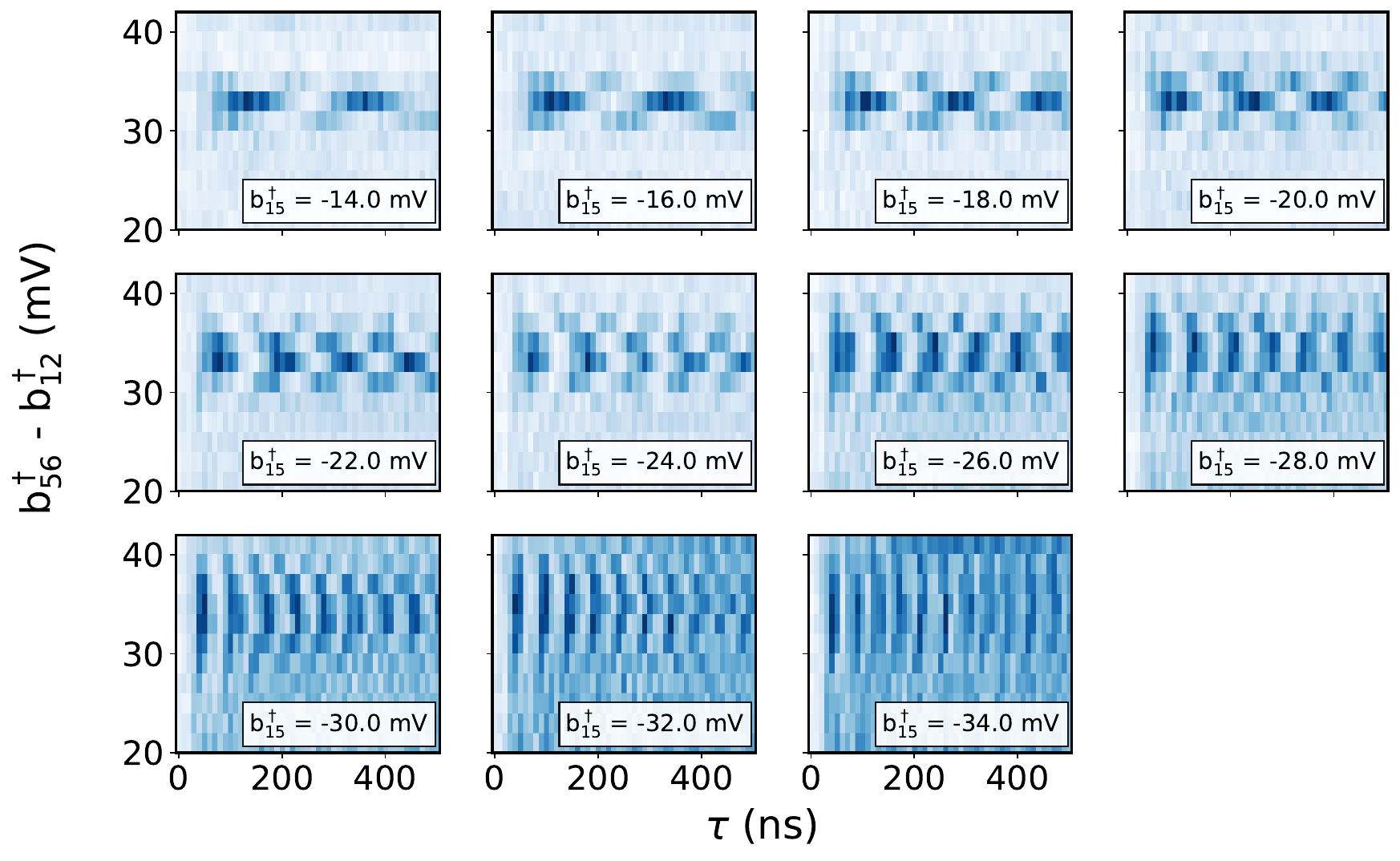}
    \caption{Resonant $\ket{S_{12}T^0_{56}}$ condition. The color scale reflects the return probability $P_{S}^{12}$ like in Fig. \ref{fig:ResonantConditions}b. We scan $b^\dagger_{56}$ from -7 to +3 mV while scanning $b^\dagger_{12}$ from -27 to -37 mV and stepping $b^\dagger_{15}$. The resonant condition is marked by a maximum in oscillation amplitude and a minimum in oscillation frequency. We observe a stable position of the resonant condition as we decrease the voltage on $b^\dagger_{15}$ showing that $J_{12}$ and $J_{56}$ remain unaltered (or more specifically, that $J_{56}-J_{12}$ remains unaltered). The chevron pattern we observe becomes broader as we decrease $b^\dagger_{15}$ which is expected as the off diagonal term in the reduced 4-spin Hamiltonian increases.}
    \label{fig:1256_resCond_vs_Vvb15}
\end{figure*}

Fig. \ref{fig:1256_resCond_vs_Vvb15} shows additional measurements of the $S_{12}T^-_{56}$ resonant condition as we step $b^\dagger_{15}$. We notice the maximum of the oscillation amplitude always in the same point for $b^\dagger_{56}-b^\dagger_{12}$ which suggests that we have likely compensated the cross-talk to $b^\dagger_{15}$ correctly. The chevron pattern becomes more spread as a result of an increase in $J_{15}$ which is reflected also by the minimum oscillation frequency in each plot. 

\subsection{Simulations of the Heisenberg chain}
\label{sec:Sims}
\begin{figure*}[h]
    \centering
    \includegraphics[width=0.9\textwidth]{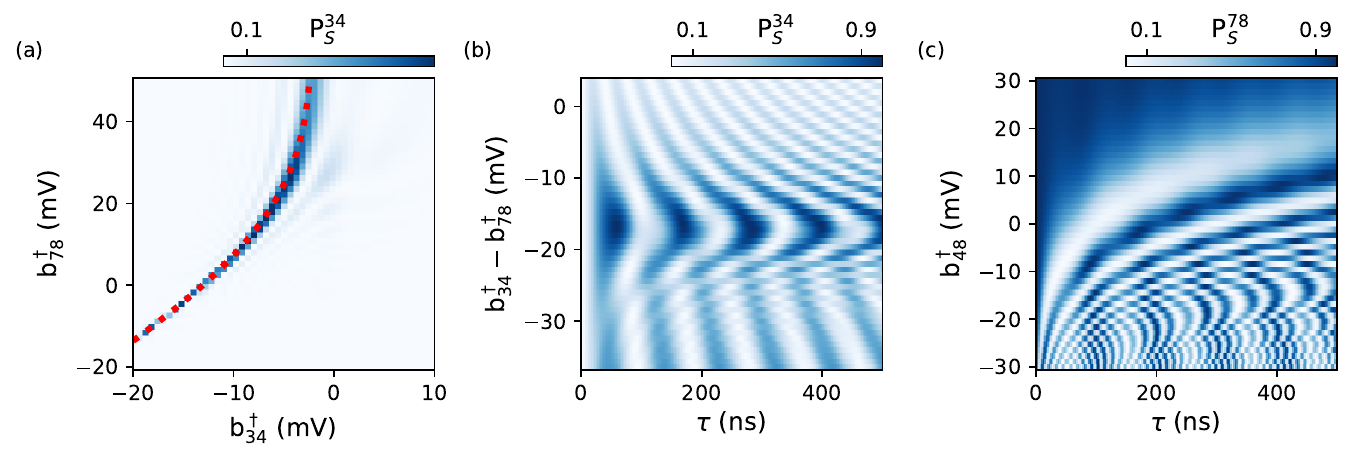}
    \caption{Simulations of the chain spanned by dots 3-4-8-7. (a) We plot the simulated singlet return probability $P_{S}^{34}$ as a function of $b^\dagger_{34}$ and $b^\dagger_{78}$ with an exchange opened between spins 4 and 8. The red dotted line is the same as in Fig. \ref{fig:ResonantConditions}a of the main text and marks the condition $|J_{34}-E_{Z34}| = |J_{78}-E_{Z78}|$. We find excellent agreement with the data.
    (b) Resonant $S_{34}T^-_{78}$ condition. We sweep $b^\dagger_{34}$ and $b^\dagger_{78}$ like in the experiment in Fig. \ref{fig:ResonantConditions}b and report $P_{S}^{34}$ finding again good agreement with the data. The leakage features for low values of $b^\dagger_{34}-b^\dagger_{78}$ are not prominent in the experiment which we attribute to a low sensor contrast.
    (c) Resonant $S_{34}T^-_{78}$ oscillations as a function of $b^\dagger_{48}$ at the resonant $ST^-$ condition. This time we report $P_{S}^{78}$ as in the experiment. The simulation matches the experiment in Fig. \ref{fig:ResonantConditions}c very well.}
    \label{fig:3478_Sims}
\end{figure*}

\begin{figure*}
    \centering
    \includegraphics[width=0.9\textwidth]{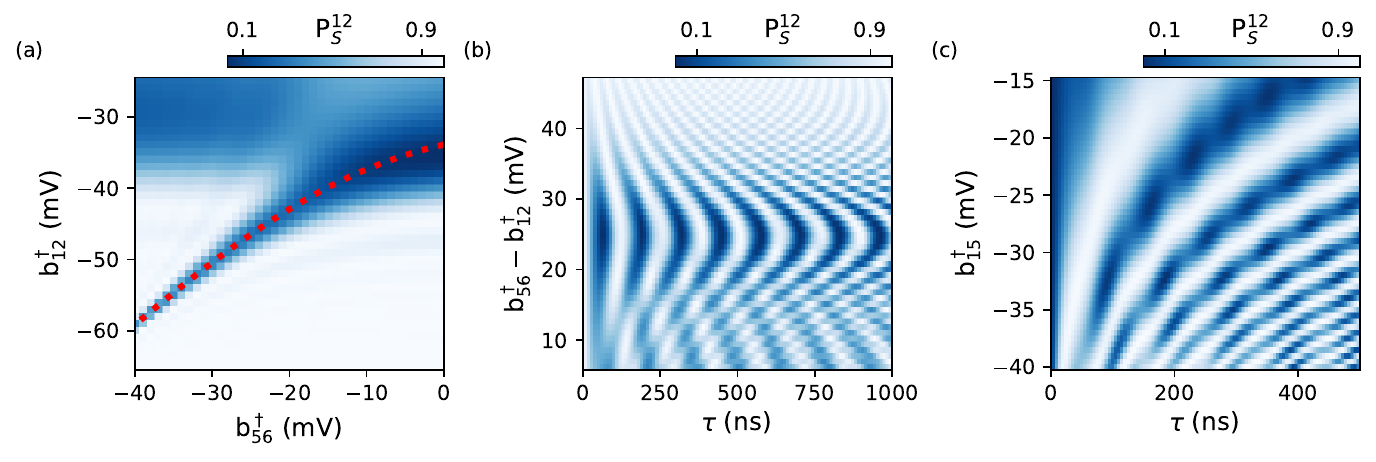}
    \caption{Simulations of the chain spanned by dots 2-1-5-6. (a) We plot the simulated singlet return probability $P_{S}^{12}$ as a function of $b^\dagger_{56}$ and $b^\dagger_{12}$ with an exchange opened between spins 1 and 5. The red dotted line is the same as in Fig. \ref{fig:ResonantConditions}e of the main text and marks the condition $|J_{12}-E_{Z12}| = |J_{56}-E_{Z56}|$. We find good agreement with the data.
    (b) Resonant $S_{12}T^-_{56}$ condition. We sweep $b^\dagger_{56}$ and $b^\dagger_{12}$ like in the experiment in Fig. \ref{fig:ResonantConditions}f and again report $P_{S}^{12}$ finding again good agreement with the data. For low values of $b^\dagger_{56}-b^\dagger_{12}$ we can also observe leakage features due to $ST^0$ oscillations in $Q_{12}$, just like in the experiment.
    (c) Resonant $S_{12}T^0_{56}$ oscillations as a function of $b^\dagger_{15}$ at the resonant $ST^0$ condition. The simulation matches the experiment in Fig. \ref{fig:ResonantConditions}g very well.}
    \label{fig:1256_Sims}
\end{figure*}

\begin{figure*}
    \centering
    \includegraphics[width=0.9\textwidth]{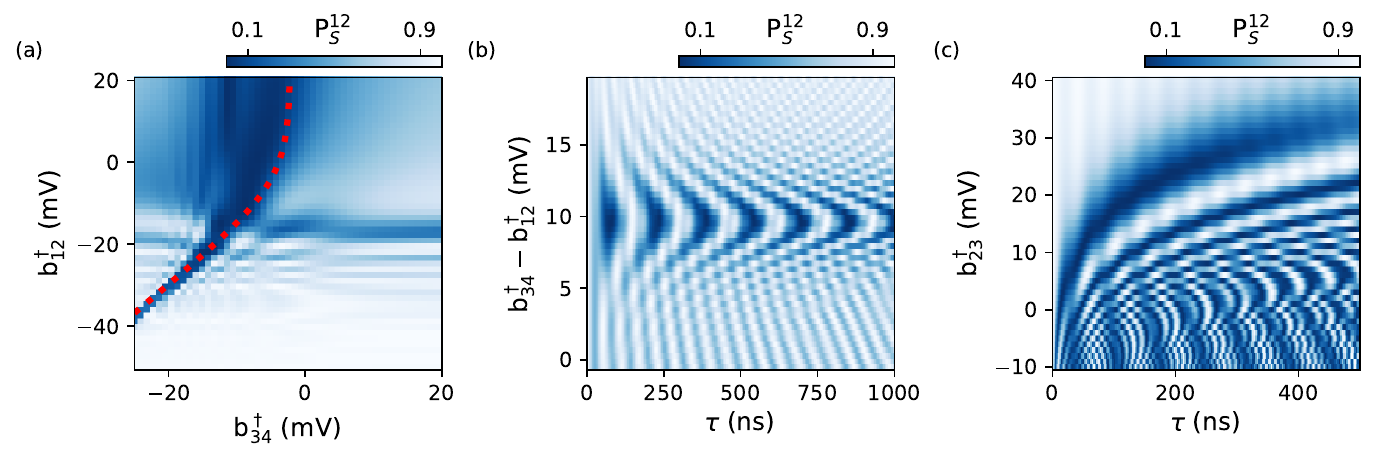}
    \caption{Simulations of the chain spanned by dots 1-2-3-4. (a) We plot the simulated singlet return probability $P_{S}^{12}$ as a function of $b^\dagger_{34}$ and $b^\dagger_{12}$ with an exchange opened between spins 2 and 3. The red dotted line is the same as in Fig. \ref{fig:ResonantConditions}i of the main text and marks the condition $|J_{34}-E_{Z34}| = |J_{12}-E_{Z12}|$. We find good agreement with the data.
    (b) Resonant $S_{12}T^-_{34}$ condition. We sweep $b^\dagger_{34}$ and $b^\dagger_{12}$ like in the experiment in Fig. \ref{fig:ResonantConditions}j and report $P_{S}^{12}$ finding again good agreement with the data except for the leakage features which are not present in the data, probably due to low visibility. 
    (c) Resonant $S_{12}T_-^{34}$ oscillations as a function of $b^\dagger_{23}$ at the resonant $ST^-$ condition. The simulation matches the experiment in Fig. \ref{fig:ResonantConditions}k well.}
    \label{fig:1234_Sims}
\end{figure*}

We perform the simulations of our experiments with the python package Qutip and assume that the system evolves under the Heisenberg Hamiltonian defined in eq. \ref{HQ}. We incorporate the experimentally determined g-factors and exchange profiles. For the 4-spin chains we operate far away from the spin-orbit avoided crossing and therefore omit them in the simulations for simplicity. Our model still captures the relevant parts of the system dynamics. Furthermore, in our analysis we have ignored g-factor modulations due to barrier gate voltages. In fact, for high enough exchange, a small deviation of the g-factors will not alter the dynamics of the system considerably. Fig. \ref{fig:3478_Sims} simulates the results for Fig. 4a,b, and c, Fig. \ref{fig:1256_Sims} pertains to the experiments in Fig. 4e,f, and g, while Fig. \ref{fig:1234_Sims} simulates the results for the linear chain obtained in Fig. 4,i,j, and k of the main text. In these three figures the red dashed lines are the same as in the corresponding figure of the main text. 

\bibliography{bibliography}

\end{document}